\documentclass[iop]{emulateapj}
\usepackage{apjfonts} 
\usepackage{graphicx}

\usepackage{amsmath,amssymb,graphicx,hyperref,twoopt,lineno,color,soul}
\usepackage{natbib}
\usepackage{threeparttable,morefloats}

\def\nms{\mathsurround=0pt}
\def\gtsim{\mathrel{\mathpalette\oversim>}} 
\def\ltsim{\mathrel{\mathpalette\oversim<}} 
\def\oversim#1#2{\lower 2pt\vbox{\baselineskip 0pt \lineskip 1pt
    \ialign{$\nms#1\hfil##\hfil$\crcr#2\crcr\sim\crcr}}}

\shorttitle{Alfv\'en Waves in BL Lacertae}
\shortauthors{Cohen et al.}

\makeindex
\citeindextrue

\received{September 10, 2014}
\accepted{January 15, 2015}
\journalinfo{Astrophysical~journal}
\submitted{}

\begin{document}
\def\deg{\ifmmode^\circ\else$^\circ$\fi}

\title{Studies of the Jet in BL Lacertae. II. Superluminal Alfv\'en Waves}

\author{
M.~H. Cohen\altaffilmark{1}, D.~L. Meier\altaffilmark{1,2},
T.~G. Arshakian\altaffilmark{3,4}, E. Clausen-Brown\altaffilmark{5}
D.~C. Homan\altaffilmark{6}, T. Hovatta\altaffilmark{7,1},
Y.~Y. Kovalev\altaffilmark{8,5}, \\
M.L. Lister\altaffilmark{9},
A.~B. Pushkarev\altaffilmark{10,11,5}, J.L. Richards\altaffilmark{9},
\and  T. Savolainen\altaffilmark{5,7}
}

\altaffiltext{1}{Department of Astronomy, California Institute of
Technology, Pasadena, CA 91125, USA; mhc@astro.caltech.edu}

\altaffiltext{2}{Jet Propulsion Laboratory, California Institute of 
Technology, Pasadena, CA 91109 USA} 

\altaffiltext{3}{I. Physikalisches Institut, Universit\"at zu K\"oln,
Z\"ulpicher Strasse 77, 50937 K\"oln, Germany}

\altaffiltext{4}{Byurakan Astrophysical Observatory, Byurakan  378433,
Armenia and Isaac Newton Institute of Chile, Armenian Branch}

\altaffiltext{5}{Max-Planck-Institut f\"ur Radioastronomie, Auf dem
H\"ugel 69, 53121 Bonn, Germany}

\altaffiltext{6}{Department of Physics, Denison University, Granville,
OH 43023 USA}

\altaffiltext{7}{Aalto University Mets\"ahovi Radio Observatory,
Mets\"ahovintie 114, 02540 Kylm\"al\"a, Finland}

\altaffiltext{8}{Astro Space Center of Lebedev Physical Institute,
Profsoyuznaya 84/32, 117997 Moscow, Russia}

\altaffiltext{9}{Department of Physics, Purdue University, 525
Northwestern Avenue, West Lafayette, IN 47907, USA}

\altaffiltext{10}{Pulkovo Observatory, Pulkovskoe Chausee 65/1, 196140
St. Petersburg, Russia}

\altaffiltext{11}{Crimean Astrophysical Observatory, 98409 Nauchny,
Crimea, Russia}


\begin{abstract}

We study the kinematics of ridge lines on the pc-scale jet of the
active galactic nucleus BL Lac. We show that the ridge lines display
transverse patterns that move superluminally downstream, and that the
moving patterns are analogous to waves on a whip. Their apparent speeds
$\rm\beta_{app}$ (units of $c$) range from 3.9 to 13.5, corresponding to
$\rm \beta_{wave}^{gal}= 0.981 - 0.998$ in the galaxy frame. We show that
the magnetic field in the jet is well-ordered with a strong transverse
component, and assume that it is helical and that the transverse
patterns are Alfv\'en waves propagating downstream on the longitudinal
component of the magnetic field. The wave-induced transverse speed of
the jet is non-relativistic ($\rm\beta_{tr}^{gal} \ltsim 0.09$).  In 2010
the wave activity subsided and the jet then displayed a mild wiggle that
had a complex oscillatory behaviour.  The Alfv\'en waves appear to
be excited by changes in the position angle of the recollimation shock,
in analogy to exciting a wave on a whip by shaking the handle.  A simple
model of the system with plasma sound speed $\rm\beta_s=0.3$ and apparent
speed of a slow MHD wave $\rm\beta_{app,S}=4$ yields Lorentz factor of the
beam $\rm\Gamma_{beam} \sim 4.5$, pitch angle of the helix (in the 
beam frame) $\alpha\sim 67\deg$, Alfv\'en speed $\rm\beta_A\sim 0.64$,
and magnetosonic Mach number $\rm M_{ms}\sim 4.7$. This describes a plasma
in which the magnetic field is dominant and in a rather tight helix, and
Alfv\'en waves are responsible for the moving transverse patterns.

\end{abstract}



\section{Introduction}  
\label{sec:intro}

This is the second in a series of papers in which we study high-resolution
images of BL Lacertae made at 15 GHz with the VLBA, under the MOJAVE
program (Monitoring of Jets in Active Galactic Nuclei with VLBA
Experiments, Lister et al., 2009).   In \citet[hereafter Paper
I]{Coh14} we investigated a quasi-stationary bright radio feature
(component) in the jet located 0.26 mas from the core, (0.34 pc,
projected) and identified it as a recollimation shock (RCS).  Numerous
components appear to emanate from this shock, or pass through it. They
propagate superluminally downstream, and their tracks cluster around
an axis that connects the core and the RCS. This behavior is highly
similar to the results of numerical modeling \citep{Lin89,Mei12}, in
which MHD waves or shocks are emitted by an RCS. In the simulations,
the jet has a magnetic field that dominates the dynamics, and is in
the form of a helix with a high pitch angle, $\alpha$.  In BL Lac the
motions of the components are similar to those in the numerical models,
and in addition the Electric Vector Position Angle (EVPA) is longitudinal;
i.e., parallel to the jet axis.  For a jet dominated by helical field,
this indicates that the toroidal component is substantial ($\rm B_{\phi}
/ B_{pol} \gtsim 1$), a necessary condition for the comparison of
the observations with the numerical simulations. Hence, in Paper I,
we assumed that the superluminal components in BL Lac are compressions
in the beam established by slow-~and/or fast-~mode magnetosonic waves
or shocks traveling downstream on a helical field.

\begin{figure*}[t]
\begin{center}
\includegraphics[width=\textwidth]{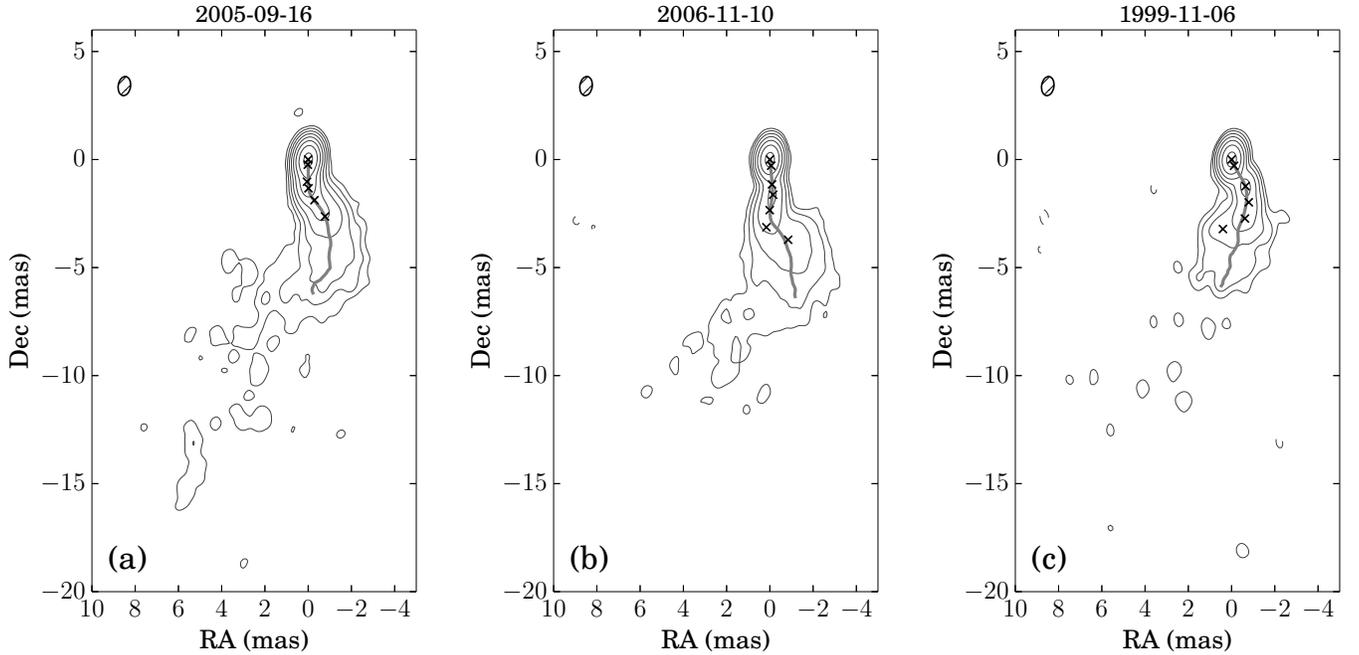}
\caption{15 GHz VLBA images of BL Lac with ridge line and components
(the crosses). In
(a) the components lie close to the ridge line. In (b) the three outer
components are off the ridge line by up to 0.3 mas.  In this case the
true ridge has a sharp bend and the algorithm has difficulty in following
it. In (c) the ridge has a step near the core, and appears to bifurcate
downstream. The algorithm misses the step, and is unable to deal with
the bifurcation.
\label{fig1}}
\end{center}
\end{figure*}

It has been common to assume that the EVPA is perpendicular to the
projection of the magnetic field vector B that is in the synchrotron
emission region. This is correct in the frame of an optically-thin
emission region, but may well be incorrect in the frame of the observer if
the beam is moving relativistically \citep{BK79, LPG05}. Lyutikov, et al.
show that if the jet is cylindrical and not resolved transversely, and if
the B field has a helical form, then the EVPA will be either longitudinal
or perpendicular to the jet, depending on the pitch angle. This is partly
seen in the polarization survey results of \citet{LH05}, where the BL
Lac objects tend to have longitudinal EVPA in the inner jet, whereas
the quasars have a broad distribution of EVPA, relative to the jet
direction. This suggests that in BL Lacs the field may be helical, with
pitch angles large enough to produce longitudinal EVPA, although strong
transverse shocks in  a largely tangled field are also a possibility
\citep[e.g.][]{H05}. The wide distribution of EVPA values in quasars
suggests that oblique shocks, rather than helical structures, might
dominate the field order. However, a distribution of helical pitch angles
could also explain the EVPAs in quasars, if symmetry is broken between
the near and far sides of the jet.  It has been suggested \citep{Mei13}
that this difference in the magnetic field is fundamental to the generic
differences between quasars and BL Lacs and, by inference, between
Fanaroff \& Riley Class II and I sources, respectively \citep{FR74}.

BL Lacs often show a bend in the jet, and the literature contains
examples showing that in some cases the EVPA stays longitudinal around
the bend; e.g., 1803+784; \citet{Gab99}, 1749+701; \citet{GP01}, and BL
Lac itself; \citet{OG09}.  In these examples the fractional polarization
$p$ rises smoothly along the jet to values as high as $p=30\%$. The field
must be well-ordered for the polarization to be that high. In this paper
we assume that the field is in a rather tight helix (in the beam frame)
and that the moving patterns (the transvese disturbances) are Alfv\'en
waves propagating along the longitudinal component of the field.

In a plasma dominated by the magnetic field, Alfv\'en waves are
transverse displacements of the field (and, perforce, of the plasma),
analogous to waves on a whip. The tension is provided by the magnetic
field $\rm (\propto B^2)$, and the wave velocity is proportional to
the square root of the tension divided by the (relativistic) mass
density. Alfv\'en waves have been employed in various astronomical
contexts, including the acceleration of cosmic rays \citep{Fer49},
the solar wind \citep{BDS69}, the Jupiter-Io system \citep{GL69},
turbulence in the ISM \citep{GS97}, the bow shock of Mars \citep{ELC10},
and the solar atmosphere\citep{Mci11}. In our case they are transverse
waves on a relativistically-moving beam of plasma threaded with a helical
magnetic field.  The appropriate formulas for the phase speeds of the
MHD waves are given in the Appendix of Paper I.

Changes in the ridge lines of BL Lacs are also seen frequently.
\citet{Bri10a} showed that in 1.4 years the BL Lac object 0735+178
changed from having a ``staircase'' structure to being straight, and that
there were prominent transverse motions.  \citet{Bri10b} also studied
1803+784 and described various models that might explain the structure.
\citet{Per12} studied the ridge line in 0836+710 at several frequencies
and over a range of epochs.  They showed that the ridge line corresponds
to the maximum pressure in the jet.  They discussed the concept of
transverse velocity, and concluded that their measured transverse motions
are likely to be caused by a ``moving wave pattern''; this was elaborated
in \citet{Per13}.  In our work here on BL Lac we also see transverse
motions, but their patterns move longitudinally and we identify them as
Alfv\'en waves.  We calculate the resulting transverse velocity of the
wave motion and show that it is non-relativistic.

It has been more customary to discuss the fast radio components in
a relativistic jet in hydrodynamic (HD) terms. We note here only a
few examples of this.  The shock-in-jet model \citep{MG85, Mar14}
was used by Hughes, Aller, \& Aller (1989a, 1989b, 1991) to develop
models of several sources, including BL Lac \citep{HAA89b} and 3C\,279
\citep{HAA91}. \citet{LZ01} recognized two threads of emission in
3C\,273 that they explained with Kelvin-Helmholtz instabilities, and
this was developed more by \citet{PLM06}.  \citet{HWG05} discussed the
patterns and motions in 3C\,120 in terms of helical instability modes.
In all these studies the magnetic field is needed of course for the
synchrotron radiation, but it also is explicitly used to explain observed
polarization changes as due to compression of the transverse components
of magnetic field by the HD shock. But in these works the magnetic
field has no dynamical role in the jet. On the contrary, in this paper,
as in Paper I, we assume that the dynamics in the jet are dominated by
the magnetic field.

\begin{figure*}[t]
\includegraphics[scale=0.9,angle=270,trim=1.8cm 0cm 5cm 0cm]{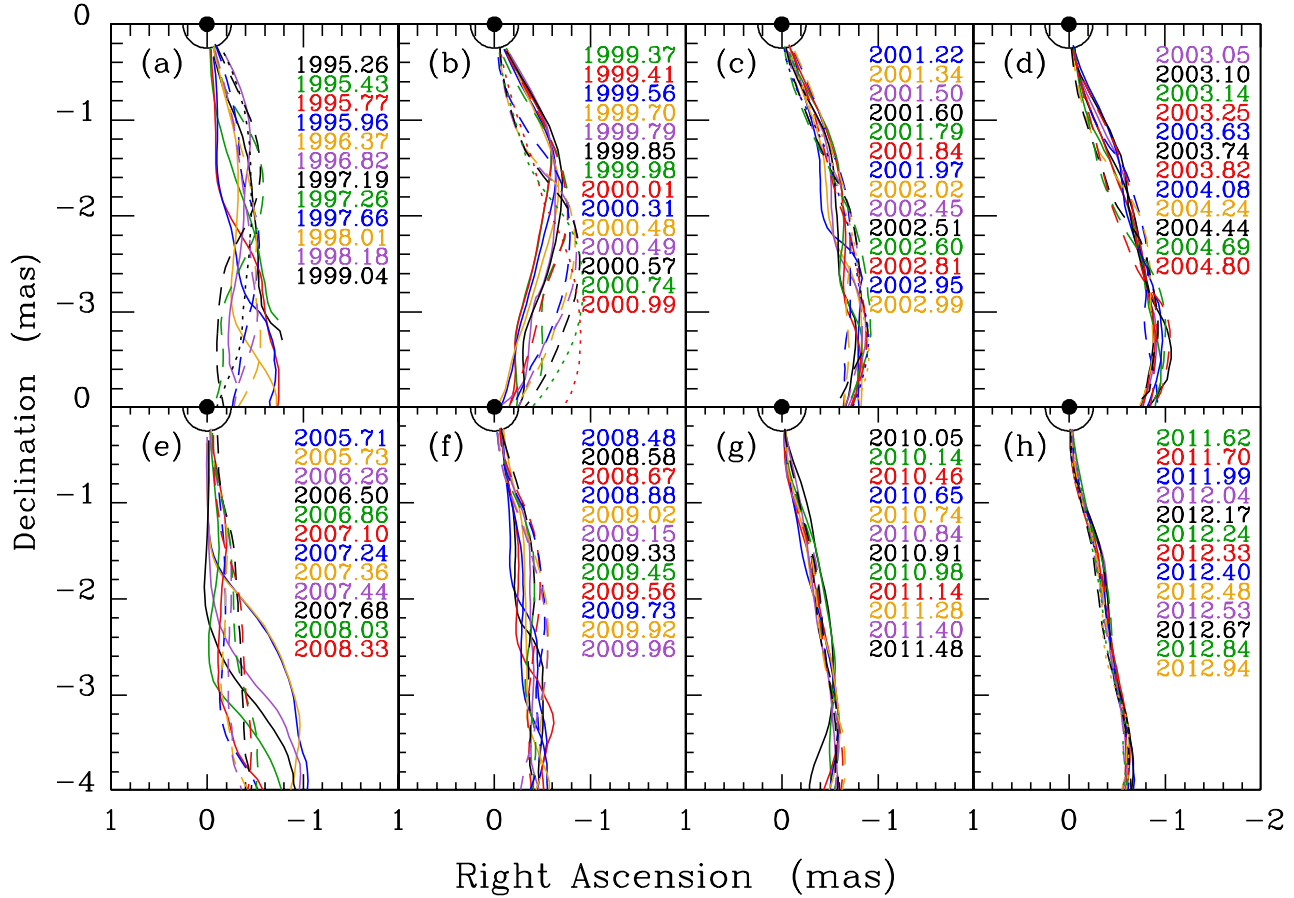}
\caption{Ridge lines for BL Lac, 1995.26 - 2012.94. Successive
panels are adjacent in time.  Epochs are identified by color. In each panel
the first occurence of a color is further identified as the solid
line, the next occurence as a dashed line, and the third occurence,
when it exists, as a dotted line. The core is shown as the solid dot,
and the semi-circle is drawn 0.25 mas from the core. In all cases the
RCS is close to the circle.
\label{fig:ridge_all}}
\end{figure*}

The plan for this paper is as follows. In Section~\ref{sec:observations}
we briefly describe the observations. The definition of the ridge line of
a jet is considered in Section~\ref{sec:ridgelines}, and the transverse
waves and their velocities, including the behavioral change in 2010,
are presented and discussed in Section~\ref{sec:waves}. Excitation
of the waves by changes in the P.A. of the RCS is considered in
Section~\ref{sec:excitation}.  In Section~\ref{Xverse_mhd} we identify
the waves as Alfv\'en waves, discuss their properties, and present 
simple models of the system.

For BL Lac $z=0.0686$, and the linear scale is $\rm 1.29~pc~mas^{-1}$.  An
apparent speed of $\rm 1~mas~yr^{-1}$ corresponds to $\rm\beta_{app}=4.20$.

\section{Observations}  
\label{sec:observations}

For this study of BL Lac we use 114 epochs of high-resolution
observations made with the VLBA at 15 GHz, between 1995.27 and
2012.98.  Most of the observations (75/114) were made under the MOJAVE
program\footnote{\url{http://www.physics.purdue.edu/astro/MOJAVE/}}
\citep{LH05}, a few were taken from our earlier 2-cm program on the VLBA
\citep{KVZ98}, and the rest were taken from the VLBA archive.

The data were all reduced by the MOJAVE team, using standard
calibration programs \citep{Lis09}. Following the reduction to fringe
visibilities we calculated three main products at nearly every epoch: 
\begin{description} 
\item[(1)]  An image, consisting of a large number of
``clean delta functions'' produced by the algorithm used for
deconvolution, convolved with a ``median restoring beam'', defined in
Section~\ref{sec:ridgelines}.
\item[(2)]  A model, consisting of a set of Gaussian
``components'' found by model-fitting in the visibility plane; each
component has a centroid, an ellipticity, a size (FWHM), and a flux
density. The Gaussians are circular when possible.  The total set of
components sums to the image, but in this paper we only use components
that have been reliably measured at four or more epochs, have flux density
$\rm >20~mJy$, and can be tracked unambiguously from epoch to epoch.
A typical epoch shows 4-6 of these ``robust'' components.  The RCS is
a permanent component and, together with the core, usually produces
more than half of the total flux density from the jet.  The centroids
of the robust components for each epoch are plotted on the images in
Figure~\ref{fig1}.

The centroid locations are measured relative to the core, which we take
to be the bright spot at the north end of the source; it usually is
regarded as the optically-thick ($\tau = 1$) region of the jet. In
principle, the core can move on the sky. We considered this in Paper I,
and concluded that any motions are less than $\rm 10~\mu as$ in a few
years, and they were ignored. Our positional accuracy is conservatively
estimated as $\rm \pm 0.1~mas$, and in this paper we again ignore any
possible core motions.
\item[(3)] The ridge line, shown in Figure 1 and discussed
in Section~\ref{sec:ridgelines}.  
\end{description} 
The image, the components, and the ridge line are not independent, but
each is advantageous when discussing different aspects of the source. In
most cases the ridge line runs down the smallest gradient from the peak
of the image, and the centroids of the components lie on the ridge
line. However, when the jet has a sharp bend the algorithm can fail, as
in Figure~\ref{fig1}c. This is discussed in Section~\ref{sec:ridgelines}.

The components move in a roughly radial direction, and plots of
$r(t)$ as well as the sky (RA--Dec) tracks are shown in Paper I and in
\citet{Lis13}.  The tracks cluster around an axis at $\mathrm{P.A.} = -166\deg$
and appear to emanate from a strong quasi-stationary component, C7, that
we identified as a recollimation shock (RCS) in Paper I.  The moving
components have superluminal speeds; the fastest has $\rm \beta_{app}
= 10\pm 1$ in units of the speed of light.  \citep{Lis13}


\section{The Ridge Lines}  
\label{sec:ridgelines}

We are dealing with moving patterns on the jet of BL Lac, and in order
to quantify them we first need to define the ridge line of a jet. At
least four definitions have been used previously.  \citet{Bri10b} used
the line that connects the components at a single epoch, in studying
1803+784.  \citet{Per12} investigated three methods of
finding the ridge line: at each radius making a transverse Gaussian fit
and connecting the maxima of the fits, using the geometrical center,
and using the line of maximum emission. They found no significant
differences among these procedures, for the case they studied, 0836+710.
They showed that the intensity ridge line is a robust structure, and
that it corresponds to the pressure maximum in the jet.

To quantify a ridge line we start with the image as in Figure~\ref{fig1},
which is the convolution of the  ``clean delta functions'' with a
smoothing beam.  Since we are comparing ridge lines from different
epochs, we have used a constant ``median beam'' for smoothing,
and not the individual (``native'') smoothing beams. The latter
vary a little according to the observing circumstances for each
epoch, and their use would effectively introduce ``instrumental
errors'' into the ridge lines.  The median beam is a Gaussian
with $\rm major~axis=0.89~mas~(FWHM),~minor~axis=0.56~ mas$ and
$\mathrm{P.A.}=-8\fdg6.$ Each of the three parameters is the median of
the corresponding parameters for all the epochs.

The algorithm for the ridge line starts at the core, and at successive
steps (0.1 mas) down the image finds the midpoint, where the integral of
the intensity across the jet, along a circular arc centered on the core,
is equal on the two sides of the arc.  The successive midpoints are then
smoothed with a third-order spline.

Ridge lines are shown on the three images in Figure~\ref{fig1}.
In Figure~\ref{fig1}a the bends in the jet are gradual and the algorithm
works very well, as indeed would any of the methods mentioned above.
In Figure~\ref{fig1}b there are two sharp bends and our algorithm makes a
smooth line that misses the corners of the bends. In this case connecting
the components would be better, if the modelling procedure actually
put components at the corners. In Figure~\ref{fig1}c the jet appears to
bifurcate, and our algorithm picks the west track. In this case a visual
inspection of the image is required to see what is going on.

In fact there is another problem with Figure~\ref{fig1}c. The image has
a step to the east (looking downstream) about 1 mas from the core, where
a short EW section connects two longer NS sections.  Since the restoring
beam is nearly NS the details of this step cannot be reconstructed. The
calculated ridge line in Figure~\ref{fig1}c does not reproduce the step,
but makes a smooth track.

\begin{figure}[b]
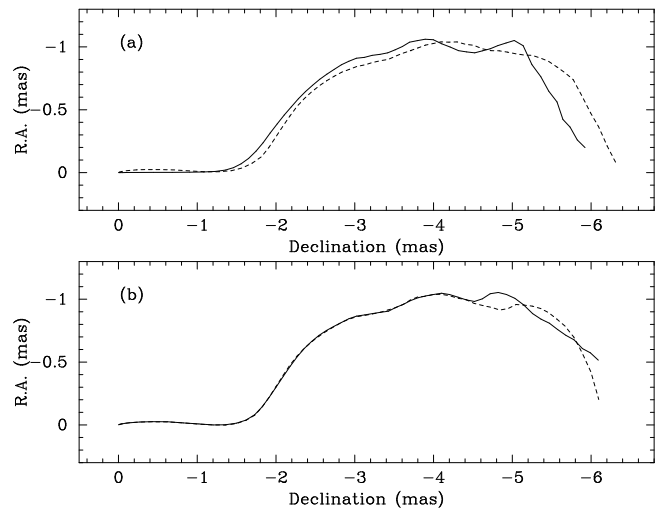

\begin{center}
\includegraphics[angle=-90,scale=0.44]{ridge_line_2005_09_16_panel_a.ps}
\includegraphics[angle=-90,scale=0.44]{ridge_line_2005_09_16_med_beam_panel_b.ps}
\caption{Ridge line for $2005-09-16$ calculated (a) with native beams 
and (b) with median beam. Solid line: using all the antennas, dotted line:
omitting SC and HN. In (a) the beam P.A.s differ by 17\deg.}
\label{fig:nativevsmedian}
\end {center}
\end {figure}

\begin{figure}[t]
\includegraphics[scale=0.89,trim=0.8cm 0.8cm 0cm 1cm]{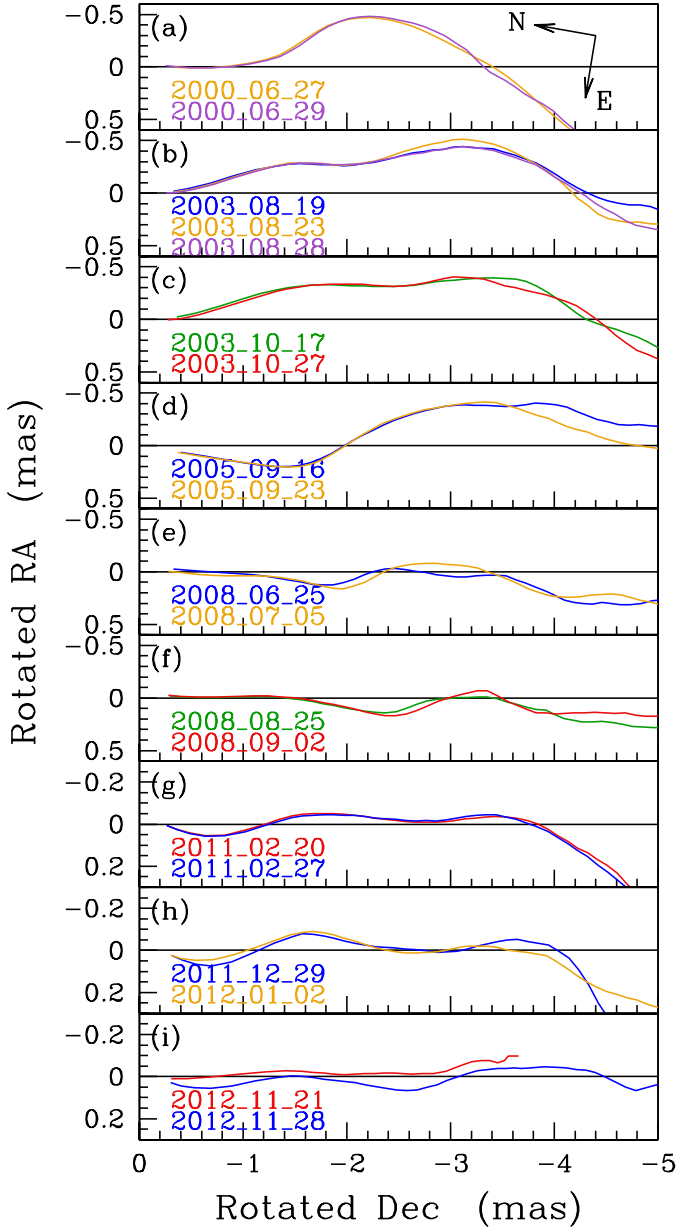}
\caption{Ridge lines for 10 pairs that each occur close in time. The axes are rotated
from (RA, Dec) by $9\fdg5$; North and East are indicated at top.
The bottom 3 panels have a different vertical scale than the others,
and the coordinate directions are thereby changed by a small amount.
\label{fig:pairs}} 
\end{figure}

Figure~\ref{fig:ridge_all} shows nearly all the ridge lines that we
consider in this paper; a few are not shown because they occur very close
in time to another one.   In all cases the RCS is located close to the
semi-circle, drawn 0.25 mas from the core.  Successive
panels are adjacent in time, although there is a 1-yr gap in the data
between panels (d) and (e). The only other substantial data gap is seen
in panel (a), from 1998.18 to 1999.04. In Figure~\ref{fig:ridge_all}
the epochs are set nearly equally among the panels, with the separations
picked to emphasize the various waves that are discussed below.

It is important to establish the reliability of the ridge lines because
our analysis rests on them, and some of the structures that we interpret
as waves are smaller than the synthesized VLBA beam.  We first note
that as with all VLBI our sampling of the $(u,v)$ plane is sparse,
and different samplings can produce different ridge lines. To see how
strong this effect is, we emulated an observation with missing antennas
by analyzing a data set with and without one and two antennas, and we
did this analysis both with the native restoring beams and the median
restoring beam described above. The results for 2005-09-16 are shown
in Figure~\ref{fig:nativevsmedian}; they are similar to the results we
obtained for two other epochs. In Figure~\ref{fig:nativevsmedian}a we show
two ridge lines, the solid one is calculated with the full data set and
the dashed line is obtained when data from the SC and HN antennas are
omitted.  The latter calculation does not use many of the baselines,
including the longest ones. The chief effect is a shift of the pattern
downstream, by roughly 0.1 mas. This shift is not a statistical effect,
but is mainly due to the different smoothing beams that were used
for the two cases.  We found that the differences in the ridge lines
increased with increasing difference in the P.A.s of the smoothing beams.
In Figure~\ref{fig:nativevsmedian}a the difference in P.A of the smoothing
beams is 17\deg.

In Figure~\ref{fig:nativevsmedian}b we used the median beam.
In this case the curves are close with differences of typically
$3~\mu$as out to 4 mas, where the surface brightness becomes low. Beyond
4 mas the differences rise to $50~\mu$as.

Another way to investigate the reliability of the ridge lines is to
examine pairs of ridge lines measured independently but close together in
time. The full data set contains 10 pairs where the separation is no more
than 10 days, and these are all shown in Figure~\ref{fig:pairs}. They are
calculated with the median restoring beam. Note that the bottom three
panels have a different vertical scale than the others.  In general
the comparison is very good within 4 mas of the core.  Panel (i)
contains one ridge line that stops at 3.6 mas because the brightness
at the ridge becomes too low; this limit also can be seen in a few
places in the other figures.  Panel (i) contains the only pair that
has a continuous offset,  $\rm 30-50~\mu as$.  These data were taken
during an exceptional flux outburst at 15 GHz in BL Lac, seen in the
MOJAVE data (unpublished), and roughly coincident with outbursts seen
at shorter wavelengths \citep{Rai13}.  An extra coreshift leading to
a position offset is expected with such an event \citep{KL08, Pus12}.
In any event, this pair appears to be different from the others, and we
do not include it in the statistics.

\begin{figure}[t]
\includegraphics[scale=0.52,trim=0cm 0.4cm 0cm 1.9cm]{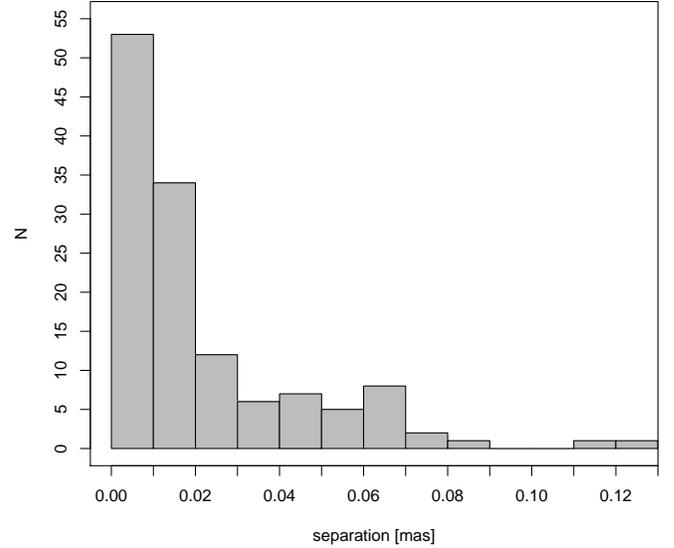}
\caption{Histogram of separations between members of 9 close
pairs of ridge lines. The pairs are shown in Figure~\ref{fig:pairs}
but panel (i) is not included in the histogram. See text.
\label{fig:histogram}}
\end{figure}

Figure~\ref{fig:histogram} shows the histogram of separations
between the paired ridge lines, after excluding those in panel (i) of
Figure~\ref{fig:pairs}.  In forming the ridge lines a 3-pixel smoothing
was used, and for the histogram we have used every third point. The
median separation is $\rm 13~\mu as$. Thus the repeatability of the
ridge lines is accurate to about $\rm 13~\mu as$. The reliability also
depends on the effect discussed in connection with Figure~\ref{fig1},
that the ridge-finding algorithm can smooth around a corner, and can
be in error by perhaps $\rm 100~\mu as$. However, the error is roughly
constant over short time spans, as in Figure~\ref{fig:pairs} panel (e)
where the sharp bend at $\rm \sim 1.5~mas$ is smoothed the same in the
two curves.  This smoothing will have little effect on calculations of
wave velocity, which is our main quantitative use of the ridge lines. We
ignore the smoothing in this paper.

From this investigation we conclude that caution must be taken in
interpreting the ridge lines, especially when comparing ridge lines
obtained at different epochs, or with different frequencies.  The details
of the restoring beam can have a noticeable effect on the ridge line,
and to avoid misinterpretation the restoring beam should be the same
for all the ridge lines that are being intercompared.

\begin{figure}[t]
\includegraphics[scale=0.9]{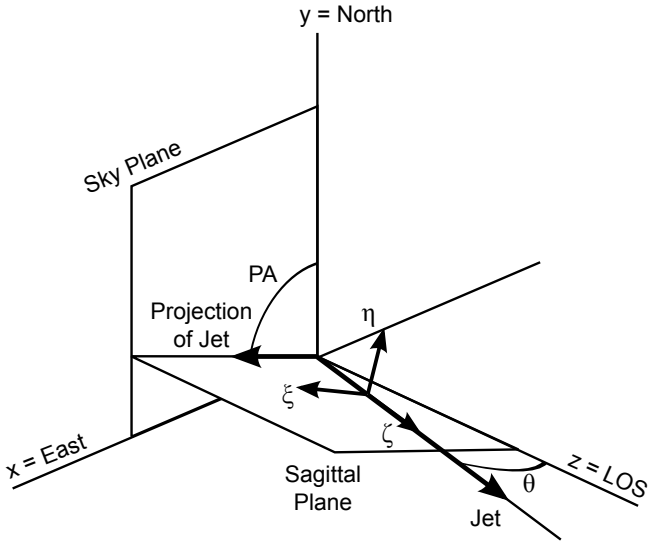}
\caption{Coordinate system. The sagittal plane is defined as the
plane containing the LOS and the mean jet axis; see text.
\label{fig:coords}}
\end{figure}

When considering these ridge lines it is important to keep the geometry
in mind: the jet has a small angle to the line-of-sight (LOS), and the
foreshortening is about a factor of 10 (Paper I). Also, the projected
images in Figure~\ref{fig1} can hide three-dimensional motions.  To work
with skew and non-planar disturbances, we use the coordinate systems
shown in Figure~\ref{fig:coords}.  East, North, and the LOS form the
left-hand system (x,y,z) and the jet lies at angle $\theta$ from the
LOS in the $sagittal~plane$\footnote{The term is taken from anatomy,
where it refers to the plane that bisects the frontal view of a figure
with bilateral symmetry. It is also used in optics, in discussions
of astigmatism.} formed by the LOS and the mean jet axis. This plane
is perpendicular to the sky plane and is at angle P.A. from the $y$
axis. The rotated system $(\xi,\eta,\zeta)$ is used to describe transverse
motions: $\xi$ is in the sagittal plane, $\eta$ is perpendicular to it,
and $\zeta$ is along the jet. By ``transverse motion'' we mean that a
point on the beam has a motion in the $(\xi, \eta)$ plane: $v_\xi,
v_\eta$. The component $v_\xi$ lies in the sagittal plane and its
projection on the sky is along the projection of the jet. This
component therefore is not visible, although
a  bright feature moving in the $\xi$ direction might be
seen as moving slowly along the jet.  However, the $v_\eta$ component
remains perpendicular to the LOS as $\theta$ or P.A. changes, and its
full magnitude is always seen. Thus a measured transverse motion is a
lower limit. If the beam is relativistic then time compression of the
forward motion must be added; see Section~\ref{sec:transverse}.

Some of the panels in Figure~\ref{fig:ridge_all} show disturbances that
appear to move down the jet, and at other epochs the jet is fairly quiet.
We now consider several of the disturbances in detail, starting with
the structures seen in Figure~\ref{fig:ridge_all}, panel (b).


\section{Waves on the Ridge Lines} 
\label{sec:waves}

\begin{figure}[t]
\includegraphics[scale=0.56,trim=0.1cm 0.4cm 0cm 1cm]{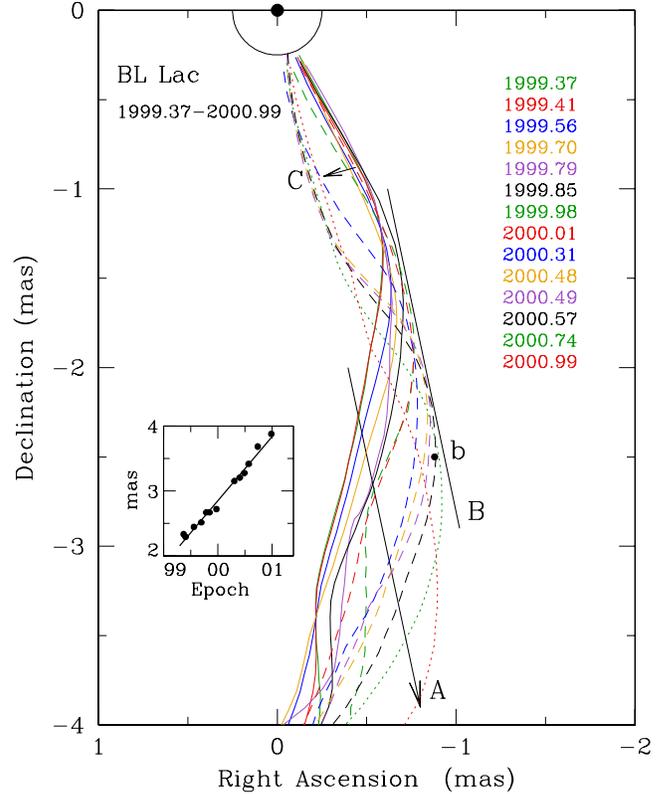} 
\caption{Ridge
lines for BL Lac at 15 GHz, for 14 epochs between 1999.37 and 2000.99.
Below $r = -2$ mas, the displacement in space corresponds to a
displacement in time, and the inset shows the points where the vector A
crosses the ridge lines -- the ordinate is distance along the vector A.
The velocity in the A direction is $\rm 0.92~mas~yr^{-1}~at~P.A.=-167\deg$;
the arrow itself represents the propagation vector that is derived in
the text.  The offset straight line B is parallel to the propagation
vector. It is approximately tangent to the wave crests, and so the
wave has constant amplitude as it moves to the SW. The short arrow C
shows a swing of the jet from west to east in early 2000; see text
Section~\ref{sec:excitation}. The point $b$ shows the characteristic
point on the 2000.57 line where the slope changes; see text
Section~\ref{sec:velocity}. Colors are as in Figure~\ref{fig:ridge_all}.
\label{fig:ridge9900}} 
\end{figure}

Figure~\ref{fig:ridge9900} is an expanded view of
Figure~\ref{fig:ridge_all}, panel (b). It includes ridge lines
for 14 consecutive epochs over a period of about 1.6 yr.  Beyond 1
mas the early epochs (solid lines) show the jet bending to the SE.
Later epochs show the bend farther downstream, and at 2000.31 and later
the jet bends to the SW before bending SE.  We anticipate a result from
Section~\ref{sec:velocity} and draw vector A at $\mathrm{P.A.}=-167\deg$
across the tracks. The intersections of vector A with the tracks
are shown in the inset in Figure~\ref{fig:ridge9900}.  The velocity
implied by the line in the inset is close to $\rm 1~mas~yr^{-1}$ or
$\rm \beta_{app} \approx 4$.  The pattern on the ridge line is moving
superluminally downstream at nearly constant velocity. We consider three
possible explanations for this.

{\bf 1)} We see the projection of a conical pattern due to a ballistic flow
from a swinging nozzle, like water from a hose. The argument against
this is that line B in Figure~\ref{fig:ridge9900} is parallel to vector
A and approximately tangent to the western crest; this  feature of
the ridge lines is not radial from the core as it would be if it were a
ballistic flow.  In Figure~\ref{fig:ridge_all} all the panels except (a),
(b), and (e) show clearly that the  excursions of the ridge lines are
constrained to lie in a cylinder, not a cone.

\begin{figure}[t]
\includegraphics[scale=0.51,angle=-90,trim=1.5cm 0cm 0cm 0cm,clip]{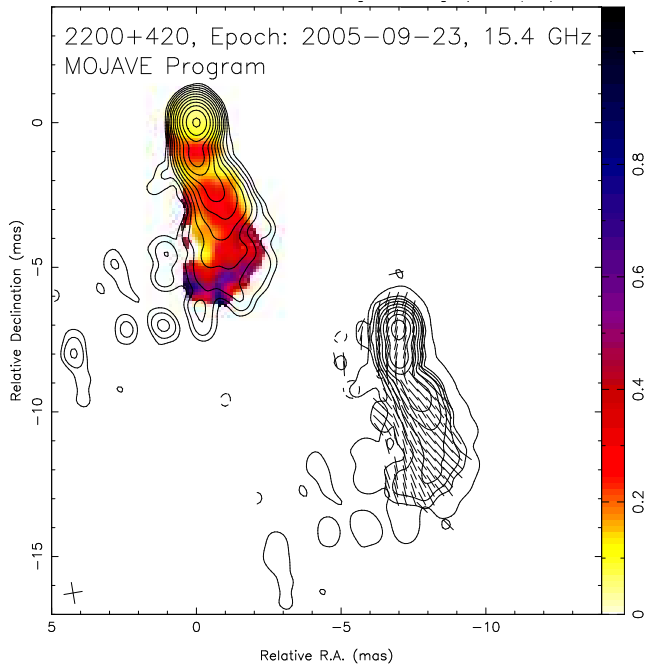}
\caption{Polarization image for BL Lac epoch 2005-09-23, one of those
forming the large wave in Figure~\ref{fig:ridge0506}.  Linear polarization
fraction $p$ is indicated by the color bar; at the core $p\approx 6\%$, in
the slice at $\sim -2~\rm{mas}~\it{p}$ drops to 15\%, and on the ridge $p$
remains near 30\% from 2 to 4 mas. In the right-hand image tick marks show
the EVPA corrected for Galactic Faraday Rotation; the EVPA stays nearly
parallel to the jet out to about 5 mas.  \label{fig:pol}} 
\end{figure}

{\bf 2)} The moving pattern is due to a helical kink instability
that is advected downstream with the flow.  In the kink the field
would be stretched out and become largely parallel to the observed bends
in the jet that, in this case, seem to be transverse waves
\citep{NM04, MHN14}.  This would produce an EVPA normal to the wave
crest in Figure~\ref{fig:ridge9900} rather than longitudinal.  But in
BL Lac the EVPA tends to be longitudinal, even along the bends.
In Figure~\ref{fig:pol} we show the polarization image for 
2005-09-23, taken from the MOJAVE website$^1$. Similar polarization
images for BL Lac, at several wavelengths, are shown in 
O'Sullivan \& Gabuzda (2009, Figure 19)
for epoch 2006-07-02. Both of these epochs are part of
the large Wave D shown later in Figure~\ref{fig:ridge0506}.  In these
polarization images the EVPA is nearly parallel to the jet out to about
5 mas and $p$ is high on the ridge, indicating that the magnetic field
remains in a relatively tightly-coiled helix around the bend and is
not nearly parallel to the axis, as it should be for an advected kink
instability.

Wave D is the largest wave in the BL Lac data, and seems to have
the cleanest longitudinal polarization. At other epochs the EVPA tends
to be longitudinal, but can be off by up to 40\deg. We have only one
epoch of polarization data for Wave A, but that one does show an EVPA
that is tightly longitudinal in the bend.  Thus we believe that the
EVPA results preclude the identification of the structures seen in
Figure~\ref{fig:ridge9900} as due to a kink instability.

{\bf 3)} The moving patterns are transverse MHD waves; i.e., Alfv\'en
waves. For this to be possible the plasma must be dynamically dominated
by a helical magnetic field. This condition for the jet of a BL Lac
has been suggested many times; see e.g., \citet{GMC04}, \citet{Mei13}.
Note that we implicitly assumed the helical, strong-field case in
discussing the kink instability, in the preceding paragraph, and we
also assumed it in Paper I. Thus, we assume that the moving pattern
under vector A in Figure~\ref{fig:ridge9900} is an Alfv\'en wave, with
velocity $\rm \sim 1~mas~yr^{-1}$.

In Figure~\ref{fig:ridge9900} a second wave is seen between $r=1$ and
$r=2$ mas, where the ridge lines for epochs 2000.31 and later bend to
the SW.  The two waves in Figure~\ref{fig:ridge9900} can be thought of
as one wave with a crest to the west.  This wave is generated by a swing
of the nozzle to the west followed by a swing back to the east about
2 years later, as discussed below in Section~\ref{sec:excitation}. 

\begin{figure}[t]
\includegraphics[width=0.49\textwidth,trim=0.1cm 0cm 0cm 0cm]{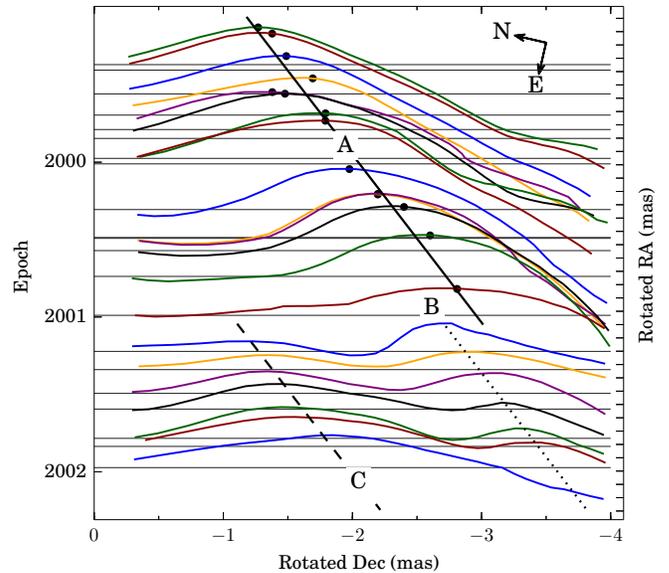}
\caption{Ridge lines for 1999.37-2001.97, plotted on axes rotated
by 13\deg. North and East are indicated at the top.
The ridge lines are spaced vertically according to epoch, and the tick
marks on the right-hand side are spaced 0.1 mas apart. The solid
line is a linear fit to the dots, which are the characteristic points
discussed in Section~\ref{sec:velocity}. The three lines are parallel and
all have slope 0.92 mas $\rm yr^{-1}$. See text.
\label{fig:waves9901}}
\end{figure}

\begin{figure}[t]
\includegraphics[scale=0.56,trim=0cm 0.3cm 0cm 0.8cm]{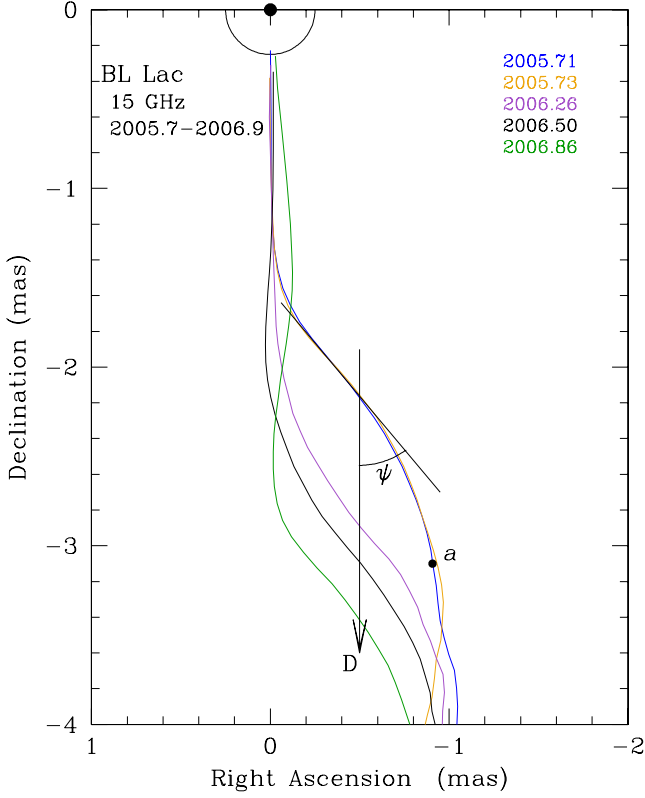}
\caption{Ridge lines for BL Lac at 15 GHz, for 5 epochs between
2005.7 and 2006.9. The propagation vector for Wave D is at $\rm{P.A.} =
-180^\circ$. 
\label{fig:ridge0506}}
\end{figure}

The 1999-2000 wave is displayed in a different form in
Figure~\ref{fig:waves9901}, which shows  the ridge lines from 1999.37 to
2001.97. Vertical spacing is proportional to epoch, and the axes have
been rotated by 13\deg; arrows at top show North and East. Tick marks
on the right are 0.1 mas apart.  The dots show the points described
later in Section~\ref{sec:velocity}, where the slope changes, and the
solid line A is a linear fit through the points, with speed $v=0.92\pm
0.05\rm{~mas~yr^{-1}}$. This wave is prominent until 2000.99. In 2001.22
the structure has changed.  There are alternate possibilities to explain
this new structure, B.  It may be a new wave, with the crests connected
with line B (drawn with the same slope as line A). In this case the
wave must have been excited somehow far from the RCS. The fit of line
B to the wave crests is poor and would be improved if acceleration
were included, but there is not enough data for that. Alternatively,
structure B may simply be a relic of the trailing side of wave A,
perhaps relativistically boosted by the changing  geometry (the bend)
seen in Figure~\ref{fig:ridge_all} panel (c).  A third wave C is shown by
the dashed line that again is drawn with the same slope.

Panel (c) of Figure~\ref{fig:ridge_all} shows the ridge lines projected
on the sky for 2001 -- 2002. Wave B from Figure~\ref{fig:waves9901}
is seen as the bump to the east at $r=2$ mas, which moves downstream
at succeeding epochs. The projected axis of the jet is curved at
these epochs, and the possible acceleration noted above for wave B may
simply be a relativistic effect inherent in the changing geometry.

Wave A in Figure~\ref{fig:waves9901} is barely visible in
Figure~\ref{fig:ridge_all} panel (a) as a gentle bump in 1999.04, so it is
first apparent in early 1999 at a distance $r\sim 1$ mas from the core.
This is reminiscent of the behavior of the components discussed in Paper
I; Figure 3 of that paper shows that most of the components first become
visible near $r=1$ mas. Wave C also appears to start near $r\sim 1$ mas.

In Figure~\ref{fig:ridge9900} the short arrow C shows an eastward
swing of the inner jet between 2000.01 and 2000.31. This is seen in
Figure~\ref{fig:waves9901} in the ridge line for 2000.31, which
shows a new inner P.A. The effect of these P.A. swings on the beam
is discussed in Section~\ref{sec:excitation}.

The different panels in Figure~\ref{fig:ridge_all} show that the jet can be
bent, and even when relatively straight, can lie at different P.A.s. Hence
there is no unique rotation angle for the ridge lines in a plot such
as that in Figure~\ref{fig:waves9901}. The rotation angle used in
Figure~\ref{fig:waves9901} was found by the velocity algorithm described
in Section~\ref{sec:velocity} for wave A. 

\begin{figure}[b]
\includegraphics[scale=0.56,trim=0cm 0.3cm 0cm 0cm]{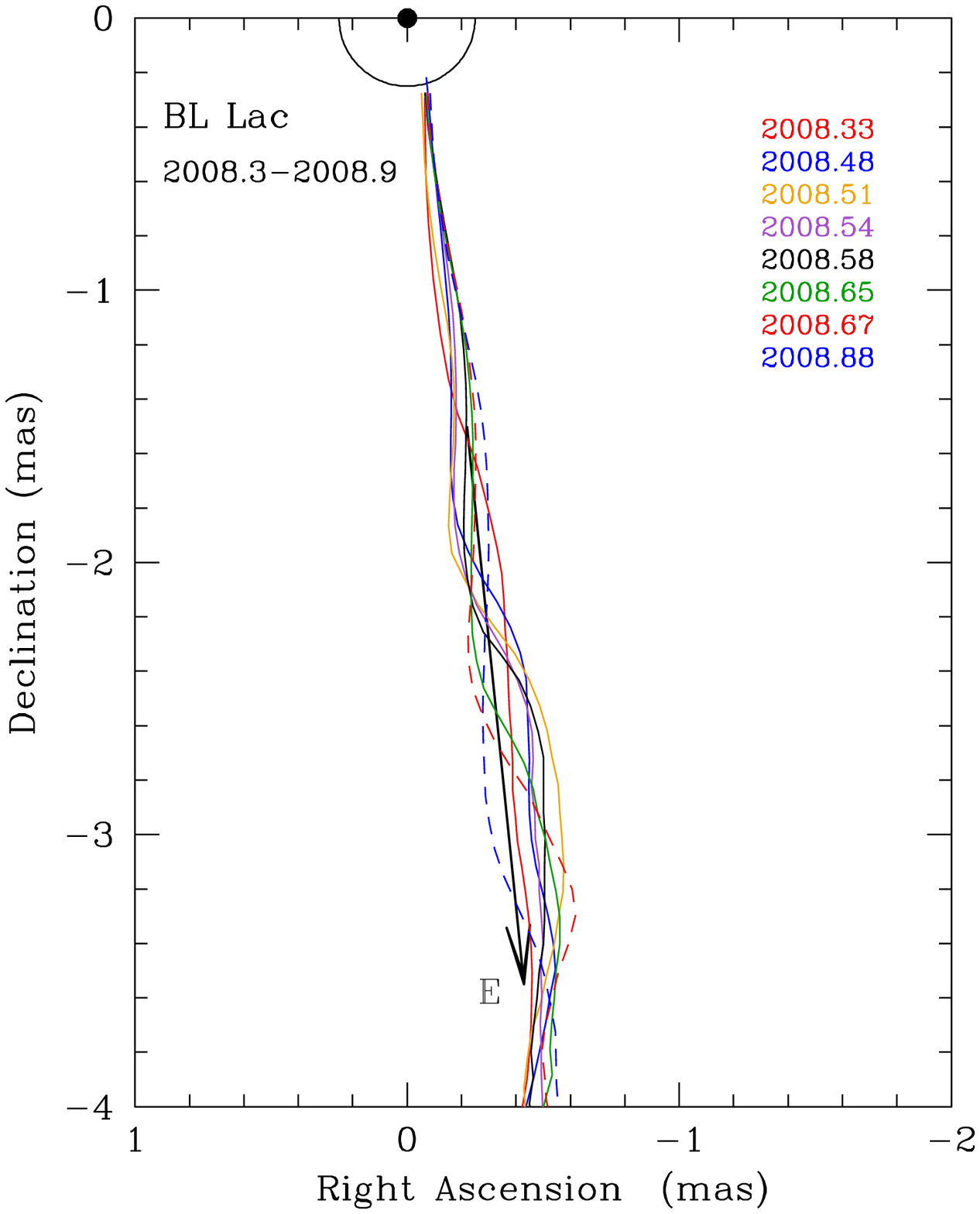}
\caption{Ridge lines for BL Lac at 15 GHz, for 7 epochs between
2008.5 and 2008.9, showing Wave E with a propagation vector 
at $\rm{P.A.} = -175^\circ$. 
\label{fig:ridge0808}}
\end{figure}

\begin{figure}[t]
\includegraphics[scale=0.56,trim=0cm 0.3cm 0cm 0.8cm]{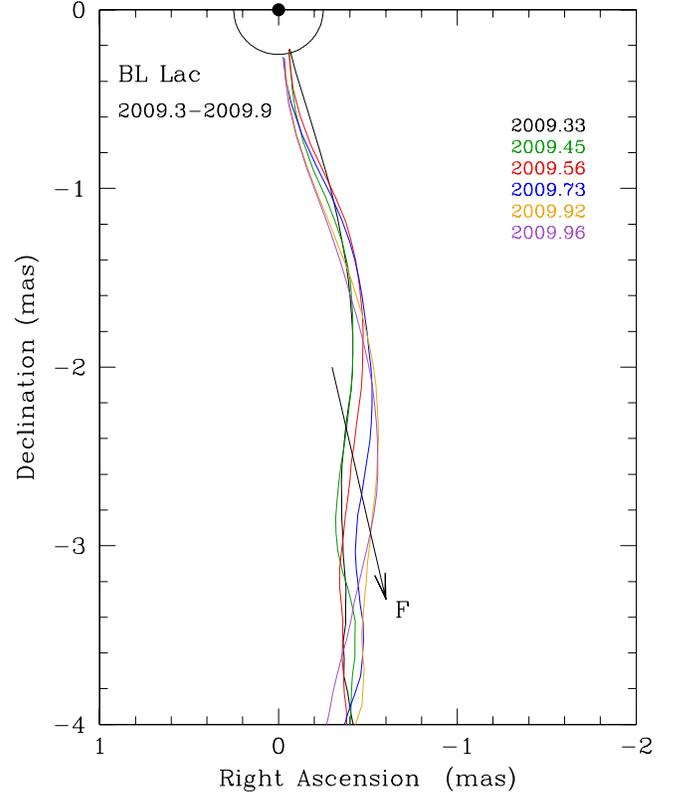}
\caption{Ridge lines for BL Lac at 15 GHz, for 6 epochs between
2009.3 and 2009.9, showing Wave F with propagation vector
at $\rm{P.A.} = -166^\circ$.
\label{fig:ridge0909}}
\end{figure}

Further examples of waves are shown in $\rm Figures~\ref{fig:ridge0506}
- \ref{fig:ridge0909}$, omitting the extraneous ridge lines to
avoid confusion.  The wave motions are indicated by the arrows,
which are propagation vectors derived in Section~\ref{sec:velocity}.
Table~\ref{tbl:waves} lists the details for these waves.  $v$ is the
measured proper motion, $\rm\beta_{app}$ is the apparent speed in units
of $c$, $\rm \beta_{wave}^{gal}$ the wave speed in the coordinate frame
of the galaxy, assuming $\theta=6\deg$, and P.A. is the  projected
direction of the propagation vector.  The amplitude is an estimate
of the projected distance (in mas) across the wave, perpendicular to the
propagation vector.  Wave D is the largest such feature seen in the
data. Unfortunately, there was an 11-month data gap prior to 2005.71,
and the wave cannot be seen at earlier times.

The amplitudes of the larger waves appear to be comparable with the
wavelength, as suggested for example by the inclination angle $\psi$
shown in Figure~\ref{fig:ridge0506}: $\psi \approx 36\deg$.  But this
is an illusion caused by the foreshortening, which is approximately a
factor of 10 (Paper I), so the deprojected value of $\psi$ is about
5\deg.  Note that this is a lower limit, since the transverse motion can
have a component in the $(\xi,\zeta)$ plane in Figure~\ref{fig:coords}.

\begin{figure}
\begin{center}
\includegraphics[width=0.47\textwidth]{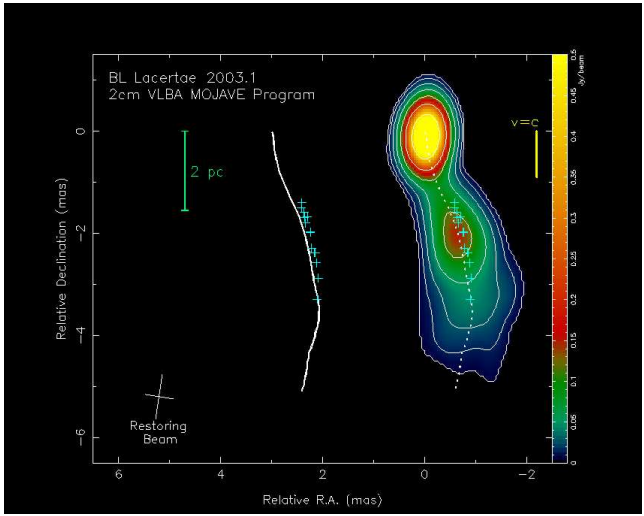}
\end{center}
\caption{Movie of the BL Lac jet at 15 GHz. The total intensity image
is on the right, with a color bar indicating flux density. The contour
levels begin at 7~mJy per beam, and increase by logarithmic factors
of 2.  The false color scheme uses a square root transfer function,
and is saturated at the core position in order to highlight changes
in the much fainter jet. The core peak brightness is highly variable;
typically it is between 2 and 6 Jy/beam.  The projected linear scale
is indicated by the 2 pc line at left.  The movie frames are linearly
interpolated between the individual VLBA epoch images, which have been
registered to the fitted position of the core feature, and restored
with a median beam with FWHM dimensions of $0.89 \times 0.57$ mas, with
a major axis position angle at $-8\fdg6$, as indicated in the lower
left corner of the frame.  The fitted ridge line is shown as a dashed
line in the image, and again as a solid line to the left of the image.
These have also been linearly interpolated between the individual VLBA
epochs.  The points of changing slope (see Section~\ref{sec:velocity})
at individual VLBA epochs are shown as the small symbols.  At left
the ridge lines are shown with different colors for the various waves.
The yellow $v=c$ line on the right is advancing at the speed of light
($\beta_\mathrm{app}=1$) and is included for reference.  The entire movie 
is available in the web version of this paper.
\label{fig:movie}}
\end{figure}

Figure~\ref{fig:movie} contains one frame of a movie of BL Lac showing
the jet motions and ridge line fits at 15~GHz. The full movie is available
in the electronic version of this paper.

\begin{table*}
\begin{center}
\caption{Transverse Waves on the Jet of BL Lac. 
\label{tbl:waves}}
\begin{tabular}{ccccccccc}
\tableline\tableline
Epoch & N & $v$ &  &$\beta_{app,T}$&$\beta^{gal}_{wave}$& P.A. && Amplitude \\
   &      & (mas y$^{-1}$)  &        &     &              &(deg) &&   (mas)   \\
\tableline
A  1999.37-2000.99&14 & 0.92 & $\pm .05$ & 3.9 & 0.979 & $-167.0$ & $\pm 1.4$ & 0.5 \\
D  2005.71-2006.86& 5 & 1.25 & $\pm .11$ & 5.6 & 0.987 & $-180.2$ & $\pm 1.1$ & 0.9 \\
E  2008.33-2008.88& 8 & 3.01 & $\pm .16$ &13.5 & 0.998 & $-174.2$ & $\pm 0.7$ & 0.3 \\
F  2009.33-2009.96& 6 & 1.11 & $\pm .19$ & 5.0 & 0.985 & $-167.1$ & $\pm 2.4$ & 0.2 \\
\tableline
\end{tabular}
\begin{tablenotes}
\small
\item
Notes. Columns are as follows:
(1) Wave label,
(2) Inclusive range of epochs,
(3) number of epochs,
(4) apparent speed,
(5) error,
(6) apparent speed in units of $c$,
(7) speed in galaxy frame, assuming $\theta=6\deg$,
(8) P.A. of the wave,
(9) error,
(10) estimated amplitude.
\end{tablenotes}
\end{center}
\end{table*}

\subsection{Different Jet Behavior in 2010-2013}
\label{sec:jet_1013}

\begin{figure}[t]
\includegraphics[scale=0.73,trim=0.2cm 0.2cm 0cm 0.2cm]{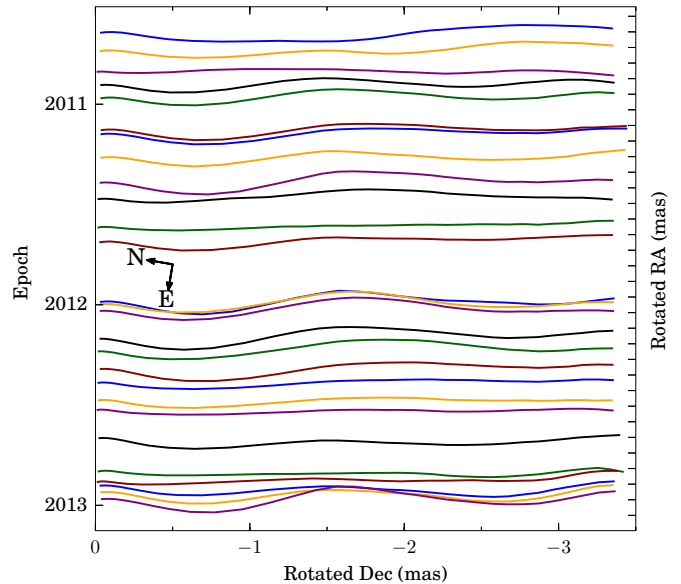} 
\caption{Ridge
lines as in Figure~\ref{fig:ridge_all} panels (g) and (h), plotted on axes
rotated by $9\fdg5$ and with vertical spacing proportional to
epoch. Tick marks on right-hand side are 0.1 mas apart.
\label{fig:ridge1013}}
\end{figure}

In Figure~\ref{fig:ridge_all} panels (g) and (h) we see that by 2010 the
earlier transverse wave activity in the jet has subsided, and that
after 2010.5 the jet is well-aligned at $\rm P.A.=-170\fdg5$ with a weak
wiggle. But the wiggle is not stationary. Figure~\ref{fig:ridge1013}
shows the ridge lines plotted on axes rotated by 9\fdg5, and spaced
proportionately to epoch. Most of the ridge lines have a
quasi-sinusoidal form.
Almost all the epochs show a negative peak in the inner jet, with a
minimum near $r=-0.7~{\rm mas}$. This is a quasi-standing feature,
of variable amplitude. At most epochs there is a positive peak near
$r = -1.6~{\rm mas}$. This also is a quasi-standing feature, but less
distinct than the inner one.

What is causing the quasi-standing features?  The patterns can hardly
be true standing waves because that requires a reflection region. A
rotating helix would project as a traveling wave, as on a barber pole,
so a simple barber-pole model is excluded.  Possible motions of the
core are only about $\rm 10~\mu as$ (Paper I), so any registration
errors due to core motion are much smaller than the observed changes,
which are up to $\rm 100~\mu as$.  There is little indication of wave
motion in Figure~\ref{fig:ridge1013}, at least not at the speeds seen
in Figure~\ref{fig:ridge_all}.  It appears then, that during the period
2010-2013, the jet was essentially straight but with a set of weak
quasi-stationary patterns, with variable amplitude.

\subsection{Velocity of the Waves}
\label{sec:velocity}

\begin{figure*}[t]
\centering
\includegraphics[scale=0.8,trim=0cm 0.5cm 0cm 0.8cm]{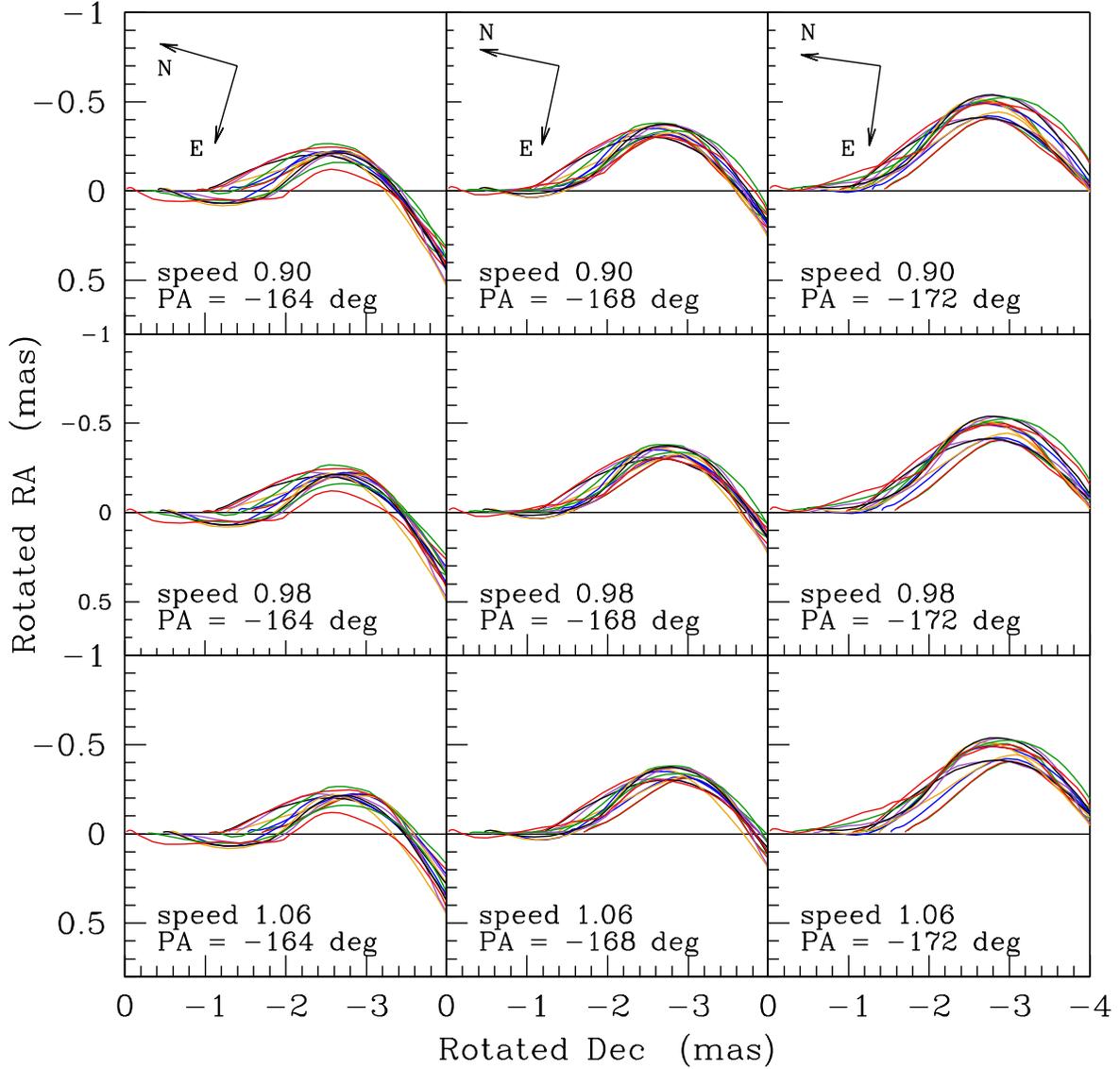}
\caption{Ridge lines shifted and overlaid on a grid of propagation
vectors, for the 14 epochs shown in Figure~\ref{fig:ridge9900}.
Each panel shows the assumed P.A. and the speed in mas s$^{-1}$; 
the P.A. is constant
in the columns and the speed is constant in the rows. 
The axes are rotated to bring the P.A. to horizontal; North and
East are shown at the top. See text.
\label{fig:superpose}}
\end{figure*}

We estimated the velocity of Wave A in Figure~\ref{fig:ridge9900} in
two independent ways.  In the first we assume that there is a constant
propagation vector, and we shift and superpose the ridge lines on a
grid of ($v$, P.A.) where $v$ is the speed of the wave and P.A. is its
propagation direction.  If the ridge lines form a simple wave, then
the solution is found when the lines lie on top of each other. This
is shown in Figure~\ref{fig:superpose}, where a reasonable fit can be
selected by eye. The result is $v=0.98 \pm 0.08~\rm{mas~yr^{-1}}$ at $\rm
P.A.=-168\deg \pm 4\deg$.  This solution is somewhat subjective and the
quoted errors do not have the usual statistical significance.

As an alternative procedure to visually aligning the ridge lines,
we developed a method of identifying a characteristic point on the
wave, just downstream of the crest, where the wave amplitude has begun
to decrease.  Define the slope of the ridge line as $\Delta x / \Delta
y$ in pixels, where in Figure~\ref{fig:waves9901}, $x$ and $y$ are
rotated RA and Dec, and take the first downstream location where the
slope exceeds $\pm 0.05$.  This point is marked with the dot $b$ on the
ridge line for 2000.57 in Figure~\ref{fig:ridge9900}.  The $x$ and $y$
positions vs time for these locations are then fit independently using
the same methods as described in \citet{Lis09} to extract a vector
proper motion for this characteristic point on the wave.

The two methods agree well and the analytic solution is $v=0.92 \pm
0.05~\rm{mas~yr^{-1}}$ at $\rm P.A.=-167\fdg0 \pm 0\fdg5$, and the
apparent speed is $\rm \beta_{app}=3.9\pm 0.2$. The propagation vector
is shown in Figure~\ref{fig:ridge9900} and the speed and
direction of the wave are listed in Table~\ref{tbl:waves}. The Table also
includes $\beta^{gal}_{wave}$ the speed of the wave in the galaxy frame,
assuming $\theta=6\deg$.  This calculation assumes that the ridge lines
lie in a plane; i.e., are not twisted.  This is not neccessarily the case.
Rather, since the inner jet, near the accretion disk, may wobble in 3
dimensions, (McKinney et al, 2013) it seems likely that the RCS may 
execute 3-dimensional motion and that the downstream jet will also.
See Section~\ref{sec:excitation}.

Note that the P.A. of the first and last propagation vectors in
Table~\ref{tbl:waves} (-167\fdg0, -167\fdg1) is the same (to within the
uncertainties) as the P.A. of the axis (-166\fdg6) defined in Paper I as
the line connecting the core with the mean position of the recollimation
shock.  In the context to be developed later, the jet acts as a whip
being shaken rapidly at the RCS, and tension in the whip continually
pulls it towards the mean PA.

In Table~\ref{tbl:waves} the speeds for the first, second and
fourth waves are all similar at $\rm \beta_{app} \sim 5$, but Wave E
(2008) is 3 times faster. Wave E has $\rm\beta_{app,E}\approx 13.5$,
which is comparable to the speed for the fastest component in BL Lac, $\rm
\beta_{app}\approx 10$, although the components speeds vary widely, from
$\rm\beta_{app} \approx 2$ to $\rm\beta_{app} \approx 10$  \citep{Lis13}.
Wave E is also distinguished by its polarization; the EVPA is transverse
not longitudinal like the others. We defer further discussion of Wave
E to another paper.

\subsection{Transverse Velocity}
\label{sec:transverse}

The ridge waves are relativistic transverse waves with apparent speeds
$\rm\beta_{app}$ from 3.9 to 13.5 times the speed of light, and we
assume that they have a small amplitude.  From the usual formula for
apparent speed,
\begin{eqnarray} 
\label{VApp} 
{\rm \beta_{app,wave}=\frac{\beta_{wave}^{gal}\sin{\theta}}
{1-\beta_{wave}^{gal}\cos{\theta}}}
\end{eqnarray} 
and taking values of $\rm\beta_{app,T}$ from Table~\ref{tbl:waves} and
using $\theta=6\deg$, we find $\beta^{gal}_{wave} = 0.979 - 0.998$ for
the speed of the waves in the frame of the host galaxy.  We now discuss
the jet motion in terms of the coordinate system $(\xi, \eta, \zeta)$
shown in Figure~\ref{fig:coords}.

Consider a transverse motion that is in the ($\eta,\zeta$) plane. Let
the beam contain a co-moving beacon that is at the origin and emits a
pulse at time $t^\prime = 0$, where $t^\prime$ is in the coordinate
frame of the galaxy.  When $t^\prime=1$ yr the signal from the
origin will have traveled $\rm 1\,ly$ down the z axis, towards the
observer. Also at $t^\prime=1$ the beacon has moved from the origin to
the point $(\eta,\zeta)=(\beta_\mathrm{tr},\beta_\mathrm{beam})$ where $\beta_\mathrm{tr}$
is the transverse speed, and $\beta_\mathrm{beam}$ is the longitudinal speed
of the beam, both in the frame of the galaxy.  At this point the beacon
emits a second signal that also travels at the speed of light.  In the
z-direction, this signal trails the first one by ($1-\beta^\mathrm{gal}_\mathrm{beam}
\cos\theta$) years. The apparent transverse speed of the beacon in
the direction perpendicular to the jet, in the galaxy frame, is then 
\begin{eqnarray} \label{Vtrans} {\beta_\mathrm{app,tr}=\frac{\beta_\mathrm{tr}}
{(1-\beta^\mathrm{gal}_\mathrm{beam}\cos\theta)}} \end{eqnarray} 
and is to be differentiated from the apparent speed $\beta_\mathrm{app}$
commonly used in studies of superluminal motion, which is the apparent
speed \textit{along} the jet.  Note the close relation
between Equations~\ref{VApp} and~\ref{Vtrans}.  Equation~\ref{Vtrans}
can be inverted to find $\rm\beta_{tr}$, a lower limit to the transverse
speed.

\begin{figure}[t]
\includegraphics[scale=0.46,trim=0.5cm 9.5cm 0cm 0.7cm]{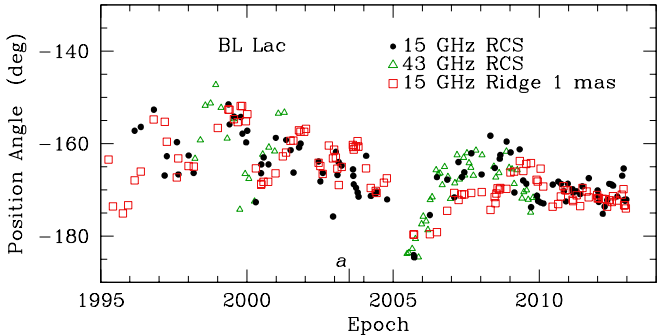}
\caption{Position Angle $vs$ Epoch for the RCS at 15 and 43 GHz,
and for the Ridge Line at $r\approx 1$ mas. Epoch {\it a} represents the 
advected start of Wave D; see text.
\label{fig:pa}}
\end{figure}

For Wave A in Figure~\ref{fig:ridge9900} we obtain an estimate
for the transverse speed at $r \sim 2$~mas by taking the
transverse motion as 0.5 mas and the time interval as $(2000.57 -
1999.41)$ yr, giving $v_{\rm tr}\approx 0.43~{\rm mas~yr}^{-1}$ and $\rm
\beta_{app,tr}=1.9$ and, from Equation~\ref{Vtrans} with $\theta=6\deg$
and $\rm\Gamma_{beam}=3.5$ (Paper I), $\beta^\mathrm{gal}_\mathrm{tr}\sim 0.09$. 
This is a model-dependent rough value, but it shows that the transverse
speed is non-relativistic. This is necessary for consistency, since
the derivation of the relativistic form of the MHD wave speeds shown in
Paper I assumes that the velocity perturbation is small.

\section{Excitation of the Waves}
\label{sec:excitation}

We suggested in Paper I that Component 7 is a recollimation shock, and
that the fast components emanate from it.  If this is correct, then the
RCS should be a nozzle and its orientation should dictate the direction of
the jet.  In this Section we investigate this possibility. We first note
that it is not possible to make a detailed mapping between the P.A. of
the RCS and the later wave shape, for two reasons.  First, the algorithm
for the ridge line smooths over 3 pixels (0.3 mas), and thus smooths
over any sharp features in the advected pattern.  The second reason is
more speculative.  Our conjecture is that the wave is launched by plasma
flowing through the nozzle and moving close to ballistically until its
direction is changed by a swing in the P.A. of the nozzle. But magnetic
tension in the jet continually pulls it towards the axis, and this means
that it will bend, and that small-scale features will be stretched out and
made smooth.

We start by comparing the P.A. of the RCS with the P.A. of the downstream
ridge line at $r=1$ mas. Figure~\ref{fig:pa} shows the P.A. of the RCS
measured at 15 GHz and at 43 GHz. The latter is calculated from data
kindly provided by the Boston University VLBI group. We used the
result found in Paper I, that the 15 GHz core is a blend of the first
two 43 GHz components and that the 15 GHz component 7 is the RCS, as is
the third 43 GHz component. We calculated the centroid of the first two
43 GHz components, to find an approximate position for the 15 GHz core,
and then calculated the P.A. of the 43 GHz RCS from that centroid. The
result is shown in Figure~\ref{fig:pa}. We eliminated one discrepant
point at 43 GHz, which was separated by about 20\deg~from nearby 43 GHz
points, and one discrepant point at 15 GHz.  The correspondence between
the two frequencies is generally good, especially after 2005.0 where the
agreement is typically within 3\deg. This further justifies our claim
(Paper I) that the the location of this component is independent of
frequency, and that it is a recollimation shock.

Figure~\ref{fig:pa} also contains the P.A. of the 15 GHz ridge line,
close to $r=$ 1.0 mas. Between 2005.0 and 2010.0 the ridge line
P.A. lags the RCS PA, by roughly 0.6 to 1.5 yr. After 2010 the PA of both
the RCS and the ridge line stabilizes, and the subsequent variations,
with rms amplitude about 3\deg, may mainly be noise.  Prior to 2005.0
the variations are faster and more frequent and the lag is erratic. In
places there appears to be no lag, but around 2000.0 and again around
2004.0 it is about 0.5 yr.  Thus it appears that the swinging in PA
of the RCS is coupled to the transverse motions of the ridge line.
When the RCS is swinging rapidly and strongly, as before 2005, then so
also is the ridge at 1 mas, with an irregular lag in P.A. that sometimes
is about a half a year, and at other times is negligible. But when
the RCS is swinging more slowly, as after 2005, then the ridge at 1 mas
is also swinging slowly, with a lag of about a year, and after 2010.0
they both are stable, with only small motions that may be dominated by
measurement errors.

We suggest that the large transverse waves on the ridge are
excited by the swinging in P.A. of the RCS. Consider Wave A, seen in
Figure~\ref{fig:ridge9900}. Its crest lies near line B and moves
downstream at $\rm 0.92~mas~yr^{-1}$. In 1999.37 the crest is at about
$r=1.2$ mas and at $\rm 0.92~mas~yr^{-1}$ would have been at the RCS
($r=0.25$ mas) around 1998.3. This is in a data gap at 15 GHz, but at
43 GHz there was a peak in P.A. in mid- or late-1998. Given that in
1999 the time lag between the RCS and the ridge at 1 mas apparently was
much less than 1 yr, the association between the peak in the RCS P.A. in
1998 and the crest of Wave A is plausible. The fall in P.A. in 1999 and
2000 is seen as the short arrow C in Figure~\ref{fig:ridge9900},
and it corresponds to the upstream side of Wave A. The downstream side
is the advected rise in P.A.  of the RCS from mid-1997 to the peak in
mid-or late 1998. The P.A. of the RCS fell from mid-1996 to mid-1997,
and we might expect to find a corresponding crest to the east on Wave A,
about 1 mas downstream of the main crest to the west.  In fact several
of the earliest ridge lines in Figure~\ref{fig:ridge9900} do show a
minor crest to the east at about $r=3.2$ mas, which is 2 mas or 2
years at 0.92 mas y$^{-1}$, downstream of the main crest to the west.
A substantial acceleration in the wave speed would be needed for this
to match.  In any event, we cannot speculate usefully on this because it
takes place beyond 3 mas, where there is a general bend to the east at
all epochs. We conclude that a plausible association can be made between
the large swing west then east of the RCS between 1998.0 and 2000.1,
and Wave A that is later seen on the ridgeline.


A similar connection can be made for Wave D, seen in
Figure~\ref{fig:ridge0506} in 2005--2006. It can plausibly be attributed
to the large swing of the RCS to the east that began in 2004 and continued
into 2005. This wave does not have a crest as Wave A does, but a crude
analysis can be made as follows.  Assume that point $a$ on the 2005.71
ridge line is the advected beginning of the wave.  With a speed of 1.25
mas y$^{-1}$ (Table~\ref{tbl:waves}) this means that the swing to the
east began around 2003.5. This date is indicated on the abscissa
in Figure~\ref{fig:pa}.  Apart from one high point at 2004.1 the P.A. of
the RCS falls gradually from 2003.1 until late 2004, when it must fall
abruptly to meet the first point after the data gap in 2005. This also
is seen in Figure~\ref{fig:ridge0506}; the first four epochs have ridge
lines that lie together and are straight at $\rm P.A.\approx -180\deg$ out
to $>1$ mas. This is consistent with the RCS P.A. being stationary
from roughly 2004.7 to 2005.7 at  $\sim -180\deg$. This is in a data
gap, and this analysis suggests that the RCS P.A. was $\approx -180\deg$
during most or all of the gap.  We conclude that the large swing
in P.A. of the RCS from mid-2003 until mid-2005 generated Wave D, the
largest wave in our data set.

The P.A. of the RCS rose rapidly from 2005.7 to about 2006.5,
but the P.A. of the ridge line rose more slowly, and not as far. From
2007.0 to 2009.0 the P.A. of the inner jet was roughly constant at about
-170\deg, while the P.A.  of the RCS slowly dropped to the same value.
We do not have a straightforward interpretation of this behavior. We also
see in Figure~\ref{fig:ridge_all} panels (e), (f) and (g) that following
the passage of Wave D the jet slowly straightened out.  The P.A. of
the inner jet $(\sim -170\deg)$ propagated as a low-amplitude wave,
at roughly the same speed as the large waves, $\rm\sim 1~mas~yr^{-1}$. 

As a further complication, during this slow straightening out
of the jet we see two more low-amplitude waves.  The high-speed Wave E
(Figure~\ref{fig:ridge0808}) has no obvious antecedent in the P.A. of
the RCS. Wave F (Figure~\ref{fig:ridge0909}) is seen a year after Wave
E, at the ``usual'' speed of 1.1 mas y$^{-1}$. These waves together
make a complex set of possibly twisted ridge lines, seen together in
Figure~\ref{fig:ridge_all} panel (f).

In Section~\ref{sec:jet_1013} we showed that the waves on the jet
subsided in 2010, and in 2010-2012 the ridge line had only a weak
variable wiggle.  During this time the P.A. of the RCS was essentially
constant; the variations seen in Figure~\ref{fig:pa} may represent the
errors in the measurements, which would be about $\pm 3\deg$. These
variations in space and time have some regularities, as discussed in
Section~\ref{sec:jet_1013}, but they do not appear to have a connection
to the P.A. of the RCS. 

In Paper I we saw that the component tracks all appear to come from or
go through the RCS (component 7) and that they lie in a window centered
on $\rm{P.A.}\approx -166\deg$ \citep{CAM12}.  This now is understood
in terms of the waves on the ridge lines, since the components all lie
on a ridge. The jet is analagous to a whip with a fixed mean axis being
shaken with small amplitudes, in various transverse directions. The whip
will occupy a narrow cylinder centered on the axis, and in projection
the cylinder becomes our window.


\section{Alfv\'en Waves and the BL Lac Whip}
\label{Xverse_mhd}

\subsection{The Transverse Waves as Alfv\'en MHD Waves Along the
Longitudinal Field Component} 
\label{Xverse_alfven}

In Paper I we showed that the magnetic field in the jet of BL Lac has
a strong transverse component. We assumed that it has a helical form,
and that it is likely that the field dominates the dynamics in the
jet. This is the condition for the existence of MHD waves that propagate
down the jet.  We suggested that the moving synchrotron-emitting
components are compressions set up by fast and/or slow magnetosonic waves,
possibly shocks. Now we introduce the third branch of MHD waves in the
jet plasma, the Alfv\'en wave, which is a transverse S (shear)
wave, with the disturbance occurring normal to the propagation direction.
In Section~\ref{sec:waves} we showed that the moving patterns on the
jet are transverse waves, and now we suggest that they are Alfv\'en waves.

The phase speed of a transverse Alfv\'en wave is given by
\begin{eqnarray} 
\label{VA} 
{\beta_{T} = \pm \beta_{A}\cos\chi}
\end{eqnarray} 
where $\beta_{A} = V_{A}/c$ is the relativistic scalar Alfv\'en speed,
given in Equation A6 of Paper I, and $\chi$ is the angle between the
propagation direction  and the magnetic field.
The Alfv\'en wave has similar propagation properties (with respect to
the magnetic field direction) as the slow wave; i.e. it moves along the
field, but not at all normal to it ($\cos\chi=0$). Note that Alfv\'en
waves generally will not produce shocks in an ideal MHD plasma.

\subsection{Calculating Physical Quantities from the Wave Speeds}
\label{subsec:physical}

We now discuss these waves in the jet and present simple models
that allow us to estimate the pitch angle $\alpha$ of the helix,
which we define as the angle between the axis of the helix and the
direction of the magnetic field when projected onto that axis.

A simple relation exists for the relativistic phase speeds of the three 
MHD waves:
\begin{eqnarray} 
\label{relation}
{\beta_\mathrm{s} = \frac{\beta_\mathrm{F}\beta_\mathrm{S}}{\beta_\mathrm{T}}}
\end{eqnarray}
where $\rm\beta_s$ is the sound speed (relative to the speed of
light), and $\rm{\beta_F, \beta_S~and~\beta_T}$ are the fast, slow,
and transverse MHD wave speeds. Equation~\ref{relation} may be readily
verified from Equation~\ref{VA} combined with Equations A1 and A2
of Paper I. With this result, the three equations for the phase
speeds, together with the definitions of the cusp and magnetosonic
speeds in Equations A3 and A4 in Paper I, can be solved for the
magnetosonic and Alfv\'en speeds: 
\begin{eqnarray} 
\label{beta_ms}
{\beta_\mathrm{ms}^2 = {\beta_\mathrm{F}^2+\beta_\mathrm{S}^2-\beta_\mathrm{F}^2\beta_\mathrm{S}^2}}
\end{eqnarray} 
\begin{eqnarray} 
\label{Alfven} 
{\beta_\mathrm{A}^2 =
\frac{\beta_\mathrm{F}^2+\beta_\mathrm{S}^2-\beta_\mathrm{F}^2\beta_\mathrm{S}^2-\beta_\mathrm{s}^2}{1-\beta_\mathrm{s}^2}}
\end{eqnarray} 
Finally, the propagation angle to the magnetic field $\chi$ can be found
from Equations~\ref{VA} and \ref{Alfven}.

In dealing with this system of equations we are helped with constraints
on the MHD wave speeds: $\beta_\mathrm{S}< \beta_\mathrm{T}< \beta_\mathrm{F} < 1$, also
$0<\beta_\mathrm{s}<1/\sqrt{3}$ for an adiabatic sound wave in a relativistic
gas.  In addition, we adopt a constraint from the one-sidedness of BL
Lac, $\Gamma_\mathrm{beam}>2.3$, where $\Gamma_\mathrm{beam}$ is the Lorentz factor
of the beam in the frame of the galaxy; this gives a jet/counterjet
intensity ratio of about $10^3$ for $\theta= 6\deg$ and a spectral
index of $-0.55$ \citep{Hov14}.  We assume that the three waves travel
downstream in the beam frame and parallel to the jet axis.  Therefore,
the propagation angle of all three waves is the pitch angle of the helix
itself: $\chi =\alpha$.

\begin{figure}[t]
\includegraphics[angle=-90,scale=0.37,trim=1cm 0cm 0.3cm 0cm]{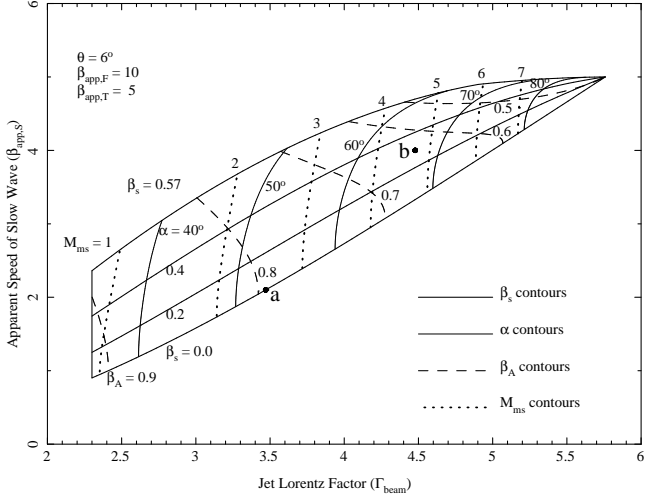}
\caption{
MHD waves along a relativistic beam containing a helical magnetic
field. The two axes show quantities defined in the galaxy frame:
jet Lorentz factor $\Gamma_\mathrm{jet}$ and the apparent (superluminal)
speed of the slow wave $\beta_\mathrm{app,S}$ (the least-known of the
three wave speeds).  The interior of the diagram contains quantities
defined in the beam frame: sound speed $\beta_\mathrm{s}$ (thick solid lines),
Alfv\'en speed $\beta_{A}$ (dashed), pitch angle of the magnetic helix
$\alpha$ (thin solid solid), and the magnetosonic Mach number $M_\mathrm{ms}$
(dotted). The region is bounded, approximately, by limits to the sound
speed, 0 and $1/\sqrt 3$, and by the limit to the Lorentz factor of the
jet, $\rm\Gamma_{beam}>2.3$, set by the limit to the jet/counterjet ratio.
The location of the diagram on the $\Gamma_\mathrm{jet}-\beta_\mathrm{app,S}$
plane depends on the values of the other observer-related
quantities, shown at upper left. The two dots show the positions of the
models discussed in the text: (a) the cold plasma ($\beta_\mathrm{s} = 0$)
model and (b) the hot plasma ($\beta_\mathrm{s} = 0.3$) model.
\label{fig:banana}} 
\end{figure}

We do not, in fact, measure the wave speeds themselves but rather
their apparent speeds in the frame of the galaxy. To relate these to
their speeds in the beam frame we first use Equation~\ref{VApp} and then the
relativistic subtraction formula
\begin{eqnarray} 
\label{beta_sub}
\beta_{{\rm wave}}^{\rm beam} & = & \frac{\beta_{{\rm wave}}^{{\rm gal}}-\beta_{\rm beam}^{{\rm gal}}}
{1  - \beta_{{\rm wave}}^{{\rm gal}} \beta_{\rm beam}^{{\rm gal}}} 
\end{eqnarray} 
where the superscripts define the coordinate frame.

We now have 5 input quantities to the calculation: $\beta_\mathrm{app,F},
\beta_\mathrm{app,S}, \beta_\mathrm{app,T}$, $\theta$ and $\Gamma_\mathrm{beam}$,
and with them we can calculate $\beta_\mathrm{s}$, $\beta_\mathrm{ms}$, $\beta_\mathrm{A}$,
$\alpha$, and the magnetosonic Mach number defined as $M_\mathrm{ms}=U_\mathrm{beam}
/ U_\mathrm{ms} = (\Gamma_\mathrm{beam} \beta_\mathrm{beam}) / (\Gamma_\mathrm{ms} \beta_\mathrm{ms})$,
where $U = \Gamma \beta$ is the magnitude of the spatial component of
the four-velocity and $\Gamma=(1-\beta^2)^{-1/2}$ is the Lorentz factor.

To illustrate the relationships among the various waves we show in
Figure~\ref{fig:banana} (the ``banana diagram'') the results for
the specific configuration $\theta=6\deg, ~\beta_\mathrm{app,F}=10,
~\beta_\mathrm{app,T}=5$. These values correspond to the fastest
superluminal component in BL Lac (Paper I) and to the apparent speeds of
the transverse waves noted in Section~\ref{sec:waves} above. The diagram
contains quantities defined in the frame of the beam: sound speed and
Alfv\'en speed, $\alpha$ the pitch angle of the helix, and $M_\mathrm{ms}$ the
magnetosonic Mach number. The diagram is bounded at the left and bottom by
$\Gamma_\mathrm{beam}=2.3$ and $\beta_\mathrm{s}=0$.  At the top, for $\alpha \lesssim
60\deg$, the boundary traces the curve $\beta_\mathrm{s} =1/\sqrt{3}$, but for
$\alpha \gtrsim 60\deg$ (in this case), this curve sometimes ventures
into a region where there are no solutions for $\alpha$.  This region
can be eliminated from the banana by continuing the curve for $\alpha >
60\deg$ with one that satisfies the criterion $d \alpha / d \Gamma_\mathrm{beam}
\approx 0$ at constant $\beta_\mathrm{S}$, as we have done here.  Inside the
banana our conditions for magnetic dominance $\beta_\mathrm{A} > \beta_\mathrm{s}$
and $M_\mathrm{ms}>1$ are satisfied everywhere except in a thin quasi-horizontal
region at top right, and in a thin quasi-vertical region at left. At the
cusp at right $\alpha=90\deg$, indicating a purely toroidal field and no
propagating Alfv\'en waves, regardless of the value of $\beta_\mathrm{s}$. The
banana diagram is set on the plane defined by the Lorentz factor of
the beam and the apparent (superluminal) speed of the slow MHD wave,
both measured in the galaxy frame. The location of the banana on this
plane is set by the specific set of input parameters as on the top left.

\subsection{Simple MHD Models of the BL Lac Jet}
\label{models}

Figure~\ref{fig:banana} shows that knowing the apparent speeds of the
three MHD waves and the angle $\theta$ of the jet to the line of
sight is not enough to completely determine the jet properties.
We must either determine one more quantity or make an assumption about the
jet system.  We will make two different assumptions for the 
sound speed, each yielding a simple model. The cases are first,
a cold jet, in which the plasma sound speed is negligible; and the other
assumes that $\beta_\mathrm{s} = 0.3$.

{\bf Model (a): Cold Plasma}

In Paper I we investigated a model of the jet in which an observed
slowly moving component with $\rm \beta_{app}=2.1$ is due to a slow
magnetosonic wave whose speed, relative to the jet plasma, is negligible:
$\beta_\mathrm{S}=0$.  This means that the plasma is cold and $\beta_s=0$
(Eqn. 4).  In this case the apparent slow component speed is the
beam speed itself. With this speed for the beam, we then assumed
that a fast component was due to a fast magnetosonic wave, and,
from the observed apparent speed, we were able to deduce its speed
on the jet.  This model can be placed in Figure~\ref{fig:banana}.
The model uses $\beta_\mathrm{s}=0, ~\theta=6\deg, ~\beta_\mathrm{app,F}=10
~\rm{and}~ \beta_\mathrm{app,S}=2.1$, and is located at the dot marked
``\textbf{a}'' on the boundary of the diagram at $\Gamma_\mathrm{beam}= 3.47,
\beta_\mathrm{app,S}=2.1$.  With $\rm\Gamma_{beam}^{gal}= 3.47$ and
$\rm\beta_{app,F}^{gal}=10$, the fast pattern speed is three times
greater than the speed of the beam, when the speeds are measured by their
Lorentz factors.  Because we now also have a measurement of the apparent
transverse Alfv\'en wave propagation speed ($\beta_\mathrm{app,T} \approx
5$, a typical value from Table 1), we can extend this model to include
computation of the total Alfv\'en speed $\beta_\mathrm{A}$, the magnetosonic
speed $\beta_\mathrm{ms}$, and the magnetic field pitch angle $\alpha$.
With $\beta_\mathrm{S}^\mathrm{beam}$ negligible, in the galaxy frame we again have
$\beta_\mathrm{beam}^\mathrm{gal} = \beta_S^\mathrm{gal} = 0.958$, $\beta_\mathrm{A}^{{\rm
gal}} = \beta_\mathrm{F}^\mathrm{gal} = 0.995$, and now $\beta_\mathrm{T}^\mathrm{gal}=
0.985$. Then, using Equation~\ref{beta_sub}, these become in the frame
of the beam $\beta_\mathrm{S}^{{\rm beam}} = 0$, $\beta_\mathrm{A}^\mathrm{beam} =
\beta_\mathrm{F}^\mathrm{beam} = 0.795$, and $\beta_\mathrm{T}^\mathrm{beam}= 0.478$,
yielding $\alpha = \cos^{-1} (0.478/0.795) = 53\deg$ -- a moderate
helical magnetic field.  Since $\beta_\mathrm{ms} = \beta_\mathrm{A}$ when $\beta_\mathrm{s}
= 0$, we also can calculate the magnetosonic Mach number defined in
Equation~\ref{beta_ms}. This yields $M_\mathrm{ms}=2.5$ and qualifies this
model as a trans-magnetosonic jet.


\begin{figure*}[t]
\includegraphics[scale=0.6, angle=-90]{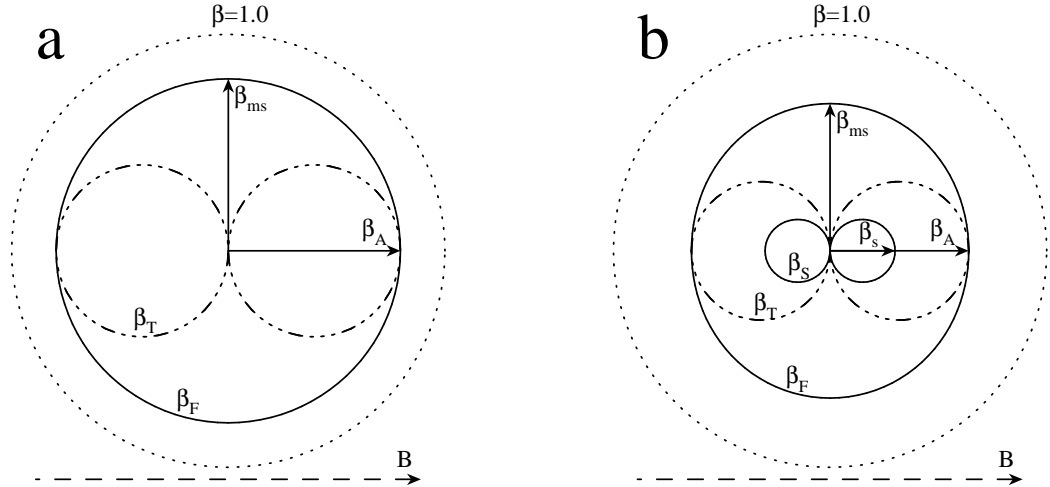}
\centering
\caption{
Relativistic phase polar diagrams for the two BL Lac jet
models discussed in the text and identified in Figure~\ref{fig:banana}.
The diagrams show the wave speed at different angles to the magnetic
field direction (dashed arrow) and are rotationally symmetric about
the horizontal direction.  (a) Model (a) (cold plasma): The slow branch
does not appear because $\beta_\mathrm{s} = 0$.  In this model $\beta_\mathrm{A} =
\beta_\mathrm{ms} = 0.794$, and the magnetic field pitch angle is $\alpha
= 53\deg$.  Having the slower Lorentz factor of the two models,
it correspondingly has the faster of the wave speeds that are consistent
with the constraints in Figure~\ref{fig:banana}.  (b) Model (b) (hot 
plasma): With $\beta_\mathrm{s} = 0.3 \neq 0$, the slow branch now appears
(innermost solid curve). Nevertheless, the magnetic field still dominates,
with $\beta_\mathrm{A} = 0.640$, $\beta_\mathrm{ms} = 0.680$, and $\alpha = 67\deg$.
\label{fig:polars}}
\end{figure*}

{\bf Model (b): Hot Plasma} 

The plasma hardly can be cold as in Model (a) because the source is
a powerful synchrotron emitter and the electron temperature is probably
of order 100 MeV; the electron component of the plasma therefore is
probably relativistic.  On the other hand, the sound speed may or may not
be near $0.577c$, depending on how heavily the plasma is contaminated
with heavy, non-relativistic ions. For lack of further information,
we choose $\beta_\mathrm{s}=0.3$. But, as seen in Figure~\ref{fig:banana}, we
still need another parameter to establish the solution. We took $\rm
\beta_\mathrm{app,S} = 2.1$ for Model (a) to match a slow superluminal component,
but if we do that now with $\beta_\mathrm{s}=0.3$ it yields $\rm\Gamma_{beam}=2.8$
and $\alpha = 43\deg$. This value for $\rm\Gamma_{beam}$ is less than that
typically found in radio beaming studies where $\rm \Gamma_{beam} \sim 7$
\citep{Jor05, Coh07, Hov09}.  Furthermore, the pitch angle $\alpha=43\deg$
is less than the one estimated from polarization analyses, $\alpha >
60\deg$. (Homan, private communication).

To reconcile these values we drop the assumption that the
superluminal component with $\rm\beta_{app}=2.1$ is a slow MHD wave
propagating downstream; it might for example be a reverse MHD shock
or wave traveling upstream in the beam frame and seen moving slowly
downstream in the galaxy frame. Instead, we choose $\rm\beta_{app,S}=4$,
because it yields acceptable values for $\rm\Gamma_{beam}$ and $\alpha$,
and matches the speed of a number of superluminal components. The final
solution, seen at point $b$ in Figure~\ref{fig:banana}, contains three
quantities that are chosen to match observations, $\theta=6\deg$,
$\rm \beta_{app,F} = 10$, $\rm \beta_{app,T} = 5$, and two quantities
picked because they are plausible and give reasonable results,
$\rm\beta_{app,S}=4$, and $\rm\beta_s = 0.3$.  The derived quantities are 
$\rm\Gamma_{beam}=4.48$,
$\rm\beta_{S}^{beam}=0.112$, 
$\rm\beta_{T}^{beam}=0.251$, 
$\rm\beta_{F}^{beam}=0.675$,
$\alpha =66\fdg9$,
$\rm\beta_A = 0.64$,
and $M_\mathrm{ms}=4.71$.  
The slow and transverse MHD waves are non-relativistic ($\beta^2\ll 1$)
in the frame of the beam, but the fast wave is mildly relativistic,
with $\rm\Gamma_F^{beam}=1.355$.  Note that relativistic addition to
produce the observed speed is non-linear; $\rm\Gamma_F^{beam}=1.355$
plus $\rm \Gamma_{beam}^{gal}=4.48$ gives $\rm \Gamma_{F}^{gal} =
10$. This is discussed in Paper~I.

With $\rm\Gamma_{beam}=4.48$ and $\theta = 6\deg$ we now calculate
the Doppler factor $\delta=7.2$, which agrees closely with values in
the literature \citep{Jor05, Hov09}. Also, $\alpha = 66\fdg9$ agrees
with estimates from polarization analyses.

Thus we see that the model for BL Lac with Alfv\'en waves on a
helical magnetic field is able to explain the moving transverse patterns
on the jet of BL Lac.  It implies a modest Lorentz factor for the actual
plasma flow ($\Gamma_\mathrm{beam} \sim 4.5$) and explains the faster propagation
of the components and the transverse disturbances as MHD sonic and
Alfv\'en waves, respectively.  They are generated primarily at the
site of the recollimation shock and propagate downstream on the helical
field, each with a speed in the galaxy frame that is the relativistic
sum of the wave speed in the beam frame and the plasma flow speed in
the galaxy frame.


The new model does, however, have a disadvantage.  The magnetosonic
Mach number of 4.7 is rather high and in conflict with the original
discussion in Paper I on the generation of a collimation shock like
C7: the super-magnetosonic flow emanating from the black hole region
should transition to a trans-magnetosonic flow ($M_\mathrm{ms} \sim 1-2$)
after it passes through C7.  However, we see in Figure~\ref{fig:banana}
that models with both low Mach number and high pitch angle are mutually
exclusive: trans-magnetosonic models have $\alpha \sim 35-47\deg$, and
models with $\alpha \sim 60-70\deg$ have magnetosonic Mach numbers of
$3.5 - 5$ or more. In order to obtain a more sophisticated model that is
compatible both with generation of a post-collimation-shock flow and the
polarization observations, it is likely that one or more of the rather
restrictive model assumptions in this paper will have to be relaxed.


Further insight into the propagation of an Alfv\'en wave on a jet can be
gained by examining the group velocity, which has only one value, $\rm
V_{A}$, and is always directed along the magnetic field \citep{GB05a}.
An isolated wave packet will spiral down the jet along the helical
magnetic field. A uniform disturbance across the jet will produce a ripple
that moves along all the field lines; i.e. across the jet. The net result
is a jump or bend that propagates downstream with speed proportional to
the cosine of the pitch angle. This has a close analogy to a transverse
mechanical wave on a coiled spring, or slinky.  In both cases there is
longitudinal tension, provided for the jet by the magnetic field.


\subsection{Phase Polar Diagrams and the Internal Properties of the 
Jet Plasma}
\label{polars}

Figure~\ref{fig:polars} shows relativistic phase polar diagrams for the
two models discussed above and identified in Figure~\ref{fig:banana}.
The diagrams show MHD wave phase speeds in 3-dimensional velocity space
with the origin of each at the center of the diagram.  Each diagram was
computed using the relativistic equations A1-A6 in Paper I.  All surfaces
are axisymmetric about the horizontal magnetic axis.  In each panel
the dotted, solid, and broken lines show respectively the speed-of-light
sphere (unity in all directions), the two compressional MHD wave surfaces
(fast [$\beta_\mathrm{F}$] and slow [$\beta_\mathrm{S}$]), and the transverse Alfv\'en
wave surface ($\beta_\mathrm{T}$). Unlike the speed of light, the speeds of
the MHD waves depend on the polar angle $\chi$ between the propagation
and field directions.  All three MHD modes are labeled in the left half
of the diagrams.  The arrows labeled in the right half of the diagrams
show the three characteristic wave speeds: sound ($\beta_\mathrm{s}$), Alfv\'en
($\beta_\mathrm{A}$), and magnetosonic ($\beta_\mathrm{ms}$), which values are realized 
along the field for the slow and Alfv\'en modes and normal to the field
for the fast mode.  As mentioned earlier, the slow and
Alfv\'en waves can propagate skew to the field, but not normal to it.

Some of the relationships among the three types of MHD waves can be seen
in Figure~\ref{fig:polars}b.  The outer solid loop traces the fast
magnetosonic mode, whose speed is a maximum $\beta_\mathrm{ms}$ at $\chi=90\deg$,
and is the same as that of the Alfv\'en wave (dashed loop), $\beta_\mathrm{A}$,
when $\chi=0\deg$, provided $\beta_\mathrm{A} > \beta_\mathrm{s}$, where $\beta_\mathrm{s}$
is the sound speed in the plasma. The propagation speed of the Alfv\'en
wave is proportional to $\cos\chi$ and this also is approximately true
for the slow magnetosonic wave, the inner loop.

\begin{figure}[t]
\centering
\includegraphics[width=0.5\textwidth,trim=4.9cm 3cm 4cm 2cm]{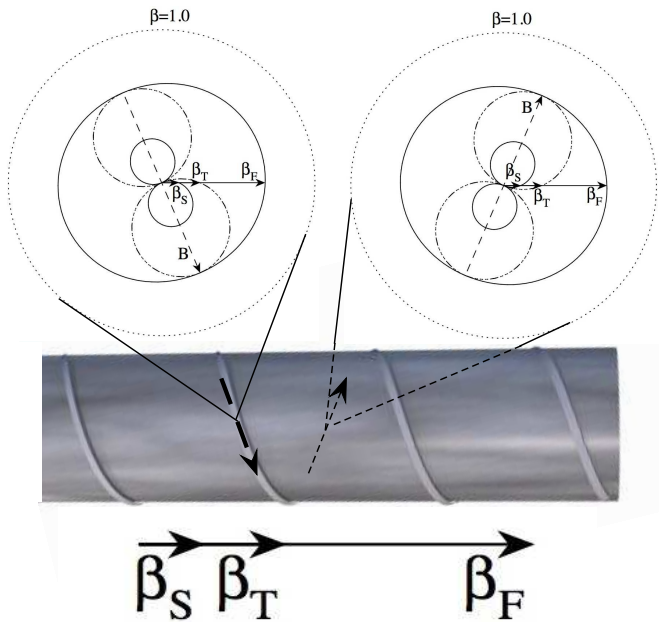}
\caption{Role of the phase polar diagram in a helical magnetic field jet
model.  We show a relativistic plasma jet (medium grey flow) in its rest
frame, described by Model (b) (discussed in the text), and wrapped with
one of its many helical magnetic field lines. At left the phase polar
diagram in Figure~\ref{fig:polars}b is rotated by the angle $-\alpha$
to align the dashed arrow with the helical field direction on the near
side of the jet (left).  The propagation speeds of the three MHD waves
along the jet axis then can be directly read off the polar diagram 
($\beta_\mathrm{S} = 0.112$, $\beta_\mathrm{T} = 0.241$, $\beta_\mathrm{F} = 0.675$).  For a
uniform helical field one obtains the same results at any point (e.g.,
on the far side of the jet at right.)
\label{fig:helicalfield}}
\end{figure}

So far we have been discussing phase polar diagrams in a
uniform magnetic field, and now address how this applies to a plasma jet
with a helical field. Figure~\ref{fig:helicalfield} shows a schematic
diagram of a helical field jet with the properties of Model (b)
discussed above and in Figures~\ref{fig:banana} and \ref{fig:polars}.
The helical field will have a pitch angle of $\alpha =\chi \approx
67\deg$, so the polar diagram in Figure~\ref{fig:polars} will be
rotated by that amount. The propagation direction of the MHD waves
points downstream in our model, allowing us to read off the values of
their propagation speeds from the polar diagram:  $\beta_\mathrm{S} = 0.112$,
$\beta_\mathrm{T} = 0.251$, and $\beta_\mathrm{F} = 0.675$. If the helical field and
plasma properties are uniform along the jet, the results will be the
same everywhere, producing MHD waves with uniform velocities.

However, there will be a longitudinal current that will cause the field
strength and pitch angle to be functions of the radial coordinate
$\varpi$. (See the cut-away view of a plasma rope in Figure 6.14
of \citet{GB05b}, for a simple view of the radial variations in B
and $\alpha$.)  But, there should be a cylindrical shell around the
axis, covering a modest range of $\varpi$, in which the synchrotron
emissivity into the direction of the observer is maximized. We assume
that this shell is the dominant region and that the field strength and
pitch angle there are the effective values that control the dynamics.
(See \citet{LPG05} for a discussion of this point. Thus, if this
is the case, then our dynamical analysis of the waves and a polarization
analysis of the emission should result in similar magnetic pitch angle
estimates for the magnetic field.  A preliminary polarization analysis
(using methods similar to those in \citet{MCG13} and to be discussed
in more detail in the next paper in this series) produces estimates of
at least 60\deg - 70\deg for the pitch angle. This is in agreement with
our result for the hot Model (b), which gives $\alpha\approx 67\deg$.

\section{Summary and Conclusions} 
\label{sec:conclusions}

The jet of BL Lac is highly variable and displays transverse patterns
that propagate superluminally downstream on the ridge line. They are not
ballistic, like water from a hose, but are constrained, like waves on
a whip.  The magnetic field is well-ordered with a strong transverse
component that we assume is the toroidal part of a helical field.
In \citet{Coh14} we assumed that the helical field provided support
for fast- and slow-mode MHD waves whose compressions we see as the
superluminal components. We here assume that the moving transverse
patterns are Alfv\'en waves propagating on the longitudinal component
of the magnetic field.

The full set of ridge lines is shown in Figure~\ref{fig:ridge_all},
and we show six examples of the Alfv\'en waves in Figures 7, 9, 10, 11,
and 12.  A movie (Figure~\ref{fig:movie}) provides assistance in studying
the motions.

The transverse wave activity died down in 2010 and the jet settled
to a fixed position angle (P.A.), with a mild wiggle. This wiggle was not
stationary, but appeared to oscillate transversely, with amplitude about
0.4 mas. This mild wiggle persisted through the remaining data period,
up to 2013.0. 

Although the transverse propagating Alfv\'en waves were greatly
reduced in 2010-2013, the superluminal components, which we identified
in Paper I as MHD sonic waves, continued roughly as before. Figure 2
in Paper I shows that during this period they continued with about the
same frequency and speed as earlier. Furthermore, during the latter half
of this period, from about 2011.4 to 2013.0, BL Lac was exceptionally
active at shorter wavelengths \citep{Rai13}, from 1 mm through gamma-rays.
This general behavior can fit into our model. We have magnetosonic
waves responsible for the superluminal components, and Alfv\'en waves
responsible for the moving transverse patterns. These are independent
MHD modes, and can be separately excited. We suspect, however, that
the increase in short-wavelength activity during the same period as the
reduction in Alfv\'en waves (2010--2013) is not a coincidence.

The velocity of the transverse waves was established by finding
characteristic points on the ridge lines where the slope changes,
as well as by visual inspection of the delayed superposition of the
ridge lines.  Three of the apparent velocities are near $\beta_{{\rm
app}}\approx 5$, and one is much faster, with $\beta_{{\rm app}}\approx
13$. With $\theta=6\deg$ and $\rm\Gamma_{beam}^{gal}=4.5$ the speeds
in the galaxy frame are approximately $\rm\beta_{T}^{gal}=0.98-0.998$
and in the beam frame $\rm\beta_{T}^{beam}=0.25-0.82$.

An Alfv\'en wave displaces the jet in the transverse direction, and
the observed motion can be converted into a transverse speed. For wave D,
the largest wave we observed, the transverse speed, in the galaxy frame,
is $\beta_\mathrm{tr}^\mathrm{gal} \sim 0.09$.  This is a rough estimate but
safely non-relativistic, and consistent with our assumption that the
waves have low amplitude.

The timing and direction of some of the the waves are correlated with
the P.A. of the recollimation shock (RCS), which swings over 25\deg~in an
irregular fashion. It appears that the waves are excited by the swinging
of the RCS. This is analogous to exciting a wave on a whip by shaking it.
In Paper I (Figure 3) we saw that the ridge lines occupy a cylinder
about 0.7 mas wide and 3 mas long, or 3 ly wide and 120 ly long when a
deprojection factor of 10 is used.  (See also \citet{CAM12} Figure 1.)
We now understand that this cylinder is formed by the transverse waves,
whose axes generally are close to the source axis at $\rm P.A. \approx
-166\deg$. The width is set by the amplitude of the largest waves while
the length is set by the general bend of the source to the SE.

We briefly describe the Alfv\'en waves, and provide a method for
calculating physical quantities in the jet in terms of the measured wave
speeds. We investigate two simple models of the system; in the first the
plasma is cold and the sound speed $\beta_\mathrm{s}=0$.  This gives results for
the Lorentz factor and the pitch angle that are in moderate disagreement
with results from observations.  The second model uses a hot plasma with
$\beta_s=0.3$, and assumes that the slow magnetosonic wave has apparent
speed $\rm\beta_{app,S}=4$.  This yields $\rm \Gamma_{beam}\approx 4.5$,
pitch angle $\alpha\approx 67\deg$, Alfv\'en speed $\rm \beta_{A}\sim
0.64$, and magnetosonic Mach number $\rm M_{ms}\approx 4.7$.  This
describes a plasma in which the helical magnetic field is strong with a
dominant toroidal component.

In our model the Lorentz factor for the beam is approximately 4.5 and
is smaller than the observed apparent speed of most of the transverse
waves as well as the fast superluminal components discussed in Paper
I. Another way to say this is that in most cases the pattern speed is
greater than the beam speed. This comes about because the pattern traces
a wave traveling downstream on the beam. 

We conclude that the rapid movements of the transverse patterns in
the jet of BL Lac can be described as Alfv\'en waves excited at the RCS
and propagating downstream on the longitudinal component of a helical
magnetic field. The jet can be described as a relativistic, rapidly
shaken whip. We suggest that other similar sources be investigated with
these ideas in mind.

\acknowledgments
We thank the referee for comments that have improved the manuscript,
M. Perucho for reading the manuscript and offering
helpful suggestions, and the MOJAVE team for comments on the
manuscript, and for years of work in producing the data base that
makes this work possible.  TGA acknowledges support by DFG project
number Os 177/2-1.  TH was partly supported by the Jenny and Antti
Wihuri foundation and by the Academy of Finland project number 267324;
TS was partly supported by the Academy of Finland project 274477.  
YYK is partly supported by the Russian Foundation for Basic Research
(project 13-02-12103), Research Program OFN-17 of the Division of
Physics, Russian Academy of Sciences, and the Dynasty Foundation.
ABP was supported by the ``Non-stationary processes in the Universe''
Program of the Presidium of the Russian Academy of Sciences.  The VLBA
is a facility of the National Radio Astronomy Observatory, a facility
of the National Science Foundation that is operated under cooperative
agreement with Associated Universities, Inc.  The MOJAVE program is
supported under NASA-Fermi grant NNX12A087G.  This study makes use of 43
GHz VLBA data from the VLBA-BU Blazar Monitoring Program (VLBA-BU-BLAZAR;
http://www.bu.edu/blazars/VLBAproject.html), funded by NASA through the
Fermi Guest Investigator Program.  Part of this research was carried out
at the Jet Propulsion Laboratory, California Institute of Technology,
under contract with the National Aeronautics and Space Administration.
This research has made use of NASA's Astrophysics Data System.


\begin{thebibliography}{}

\bibitem[Belcher, Davis \& Smith(1969)]{BDS69}
Belcher, J.~W., Davis, L. Jr. \& Smith, E.~J. 1969, \jgr, 74, 2302


\bibitem[Blandford \& K\"onigl(1979)]{BK79}
Blandford, R.~D. \& K\"onigl, A. 1979, \apj, 232, 34

\bibitem[Britzen et al.(2010a)]{Bri10a}
Britzen, S. et al 2010a, \aap, 515, A105 

\bibitem[Britzen et al.(2010b)]{Bri10b}
Britzen, S. et al 2010b, \aap, 511, A57

\bibitem[Caproni, Abraham, \& Monteiro(2012)]{CAM12}
Caproni, A., Abraham, Z. \& Monteiro, H. 2012, \mnras, 428, 280

\bibitem[Cohen et al.(2007)]{Coh07}
Cohen, M.~H., Lister, M.~L., Homan, D.~C., et al. 2007,
\apj, 658, 232

\bibitem[Cohen et al.(2014)]{Coh14}
Cohen, M.~H. et al. 2014, \apj, 787, 151 (Paper I)

\bibitem[Edberg et al.(2010)]{ELC10}
Edberg, N.~J.~T. et al. 2010, \jgr, 115, AO7203

\bibitem[Fanaroff \& Riley(1974)]{FR74}
Fanaroff, B.~L. \& Riley, J.~M. 1974, \mnras, 167, 31

\bibitem[Fermi(1949)]{Fer49}
Fermi, E. 1949, Physical Review, 75, 1169

\bibitem[Gabuzda(1999)]{Gab99}
Gabuzda, D.~C. 1999, New Astronomy Reviews, 43, 691

\bibitem[Gabuzda, Murray \& Cronin(2004)]{GMC04}
Gabuzda, D.~C., Murray, \'E. \& Cronin, P.~J. 2004, \mnras, 351, L89

\bibitem[Gabuzda \& Pushkarev(2001)]{GP01}
Gabuzda, D.~C. \& Pushkarev, A. 2001, in ASP Conf. Ser. 250, Particles 
and Fields in Radio Galaxies, eds R.~A. Liang \& K.~M. Blundell, 180

\bibitem[Goldreich \& Lynden-Bell(1969)]{GL69}
Goldreich, P. \& Lynden-Bell, D. 1969, \apj, 156, 59

\bibitem[Goldreich \& Sridhar(1997)]{GS97}
Goldreich, P., \& Sridhar, S. 1997, \apj, 438, 763

\bibitem[Gurnett \& Bhattacharjee(2005)]{GB05a}
Gurnett, D.~A. \& Bhattacharjee, A. 2005, ``Introduction to 
Plasma Physics'', Cambridge, p199

\bibitem[Gurnett \& Bhattacharjee(2005)]{GB05b}
Gurnett, D.~A. \& Bhattacharjee, A. 2005, ``Introduction to 
Plasma Physics'', Cambridge, p208

\bibitem[Hardee, Walker \& G\'omez(2005)]{HWG05}
Hardee, P.~E., Walker, R.~C. \& G\'omez, J.~L. 2005, \apj, 620, 646

\bibitem[Hovatta et al.(2009)]{Hov09}
Hovatta, T., Valtaoja, E., Tornikoski, M. \& Lahteenmaki, A. 2009, 
\aap, 494, 527

\bibitem[Hovatta et al.(2014)]{Hov14}
Hovatta, T. et al. 2014, \aj, 147, 143

\bibitem[Hughes(2005)]{H05}
Hughes, P.~A. 2005, \apj, 621, 635

\bibitem[Hughes Aller \& Aller(1989a)]{HAA89a}
Hughes, P.~A., Aller, H.~D. \& Aller, M.~F. 1989a, \apj, 341, 54

\bibitem[Hughes, Aller, \& Aller(1989b)]{HAA89b}
Hughes, P.~A. Aller, H.~D. \& Aller, M.~F. 1989b, \apj, 341, 68

\bibitem[Hughes, Aller, \& Aller(1991)]{HAA91}
Hughes, P.~A. Aller, H.~D. \& Aller, M.~F. 1991, \apj, 374, 57

\bibitem[Jorstad et al.(2005)]{Jor05}
Jorstad, S.~G. et al. 2005, \aj, 130, 1418

\bibitem[Kellermann et al.(1998)]{KVZ98}
Kellermann, K.~I., Vermeulen, R.~C., Zensus, J.~A. \& Cohen, M.~H. 1998,
\aj, 115, 1295

\bibitem[Kovalev et al.(2008)]{KL08}
Kovalev, Y.~Y., Lobanov, A.~P., Pushkarev, A.~B. \& Zensus, J.~A. 
2008, \aap, 483, 759

\bibitem[Lind et~al.(1989)]{Lin89}
Lind, K.~R., Payne, D.~G., Meier, D.~L., \& Blandford, R.~D. 1989,
\apj, 344, 89

\bibitem[Lister \& Homan(2005)]{LH05}
Lister, M.~L. \& Homan, D.~C. 2005, \aj, 130, 1389

\bibitem[Lister et al(2009)]{Lis09}
Lister, M.~L., Cohen, M.~H., Homan, D.~C., et al. 2009, \aj, 138, 1874

\bibitem[Lister et al(2013)]{Lis13}
Lister, M.~L., Aller, M.~F., Aller H.~D., et al. 2013, \aj, 146, 120

\bibitem[Lobanov \& Zensus(2001)]{LZ01}
Lobanov, A.~P. \& Zensus, J.~A. 2001, Science, 294, 128

\bibitem[Lyutikov, Pariev, \& Gabuzda(2005)]{LPG05}
Lyutikov, M., Pariev, V.~I. \& Gabuzda, D.~C. 2005, \mnras, 360, 869

\bibitem[Marscher \& Gear(1985)]{MG85}
Marscher, A.~P. \& Gear, W.~K. 1985, \apj, 298, 114

\bibitem[Marscher(2014)]{Mar14}
Marscher, A.~P. 2014, \apj, 780, 87

\bibitem[McIntosh et al.(2011)]{Mci11}
McIntosh, S.~W., de Pontieu, B., Carlsson, M., Hansteen, V., Boerner, P.
\& Goossens, M. 2011, \nat, 475, 477 

\bibitem[Meier(2012)]{Mei12}
Meier, D.~L. ``Black Hole Astrophysics: The Engine Paradigm'', Springer,
2012, 717 


\bibitem[Meier(2013)]{Mei13}
Meier, D.~L. 2013 in ``The Innermost Regions of Relativistic Jets and Their
Magnetic Fields'', EPJ Web of Conferences, 61, 01001

\bibitem[Mizuno, Hardee, \& Nishikawa(2014)]{MHN14}
Mizuno, Y., Hardee, P.~E. \& Nishikawa, K.-I. 2014, \apj, 784

\bibitem[Nakamura \& Meier(2004)]{NM04}
Nakamura, M. \& Meier, D.~L. 2004, \apj, 617, 123

\bibitem[Nakamura \& Meier(2014)]{NM14}
Nakamura, M. \& Meier, D.~L. 2014, \apj, 785, 152

\bibitem[Murphy, Cawthorne \& Gabuzda(2013)]{MCG13}
Murphy, E., Cawthorne, T.~V. \& Gabuzda, D.~C. 2013, \mnras, 430, 1504

\bibitem[O'Sullivan \& Gabuzda(2009)]{OG09}
O'Sullivan, S.~P. \& Gabuzda, D.~C. 2009a, \mnras, 393, 429.


\bibitem[Perucho et al.(2006)]{PLM06}
Perucho, M., Lobanov, A.~P., Mart\'i, J.-M., \& Hardee, P.~E. 2006, 
\aap, 456, 493

\bibitem[Perucho et al(2012)]{Per12}
Perucho, M., Kovalev, Y.~Y., Lobanov, A.~P., Hardee, P.~E. \& Agudo, I. 2012,
\apj, 749, 55

\bibitem[Perucho(2013)]{Per13}
Perucho, M.\ 2013, European Physical Journal Web of Conferences, 61, 2002

\bibitem[Pushkarev et al.(2012)]{Pus12}
Pushkarev, A.~B. et al. 2012, \aap, 545, A113

\bibitem[Raiteri et al.(2013)]{Rai13}
Raiteri, C.~M. et al. 2013, \mnras, 436, 1530

\end{thebibliography}
\end{document}